\documentclass[pre,preprint,showpacs,showkeys,amsfonts]{revtex4}

  \usepackage{latexsym}

  \usepackage{amssymb}

  \usepackage{graphicx}

  \newtheorem{lemma}{Lemma}[section]

  \newtheorem{theorem}{Theorem}[section]

\begin{document}
  \title{Modelling of quantum networks}

\author{Anna B. Mikhailova}
 \email{baltic@teia.org}
 \affiliation{Laboratory of Quantum Networks,\, V.A.Fock Institute for  Physics
 of St.Petersburg   University,\, St.Petersburg 198504, Russia}
 \vskip10pt

 \author{Boris S. Pavlov}
\email{pavlov@math.auckland.ac.nz} \affiliation{Department of
Mathematics, University of Auckland, Private Bag 92019, Auckland,
New Zealand} \vskip10pt

\author{Lev Prokhorov}
 \email{Lev.Prokhorov@pobox.spbu.ru}
 \affiliation{Laboratory of Quantum Networks,\, V.A.Fock Institute for  Physics
 of St.Petersburg   University,\, St.Petersburg 198504, Russia}
\vskip10pt

 \vskip 1truecm

            \thispagestyle{empty}
            \baselineskip=15pt
            \tolerance=600

 \begin{abstract}

\noindent {\small Mathematical design of a quantum network with
prescribed  transport properties is equivalent to the Inverse
Scattering Problem for Schr\"{o}dinger equation on a composite
domain consisting of quantum wells and finite or semi-infinite
quantum wires attached to them. An alternative approach to study
of transport properties of  networks, based on solution of the
direct Scattering problem, is also possible, but optimization of
design requires scanning  over the multi-dimensional space of
essential physical and geometrical parameters of the  network. We
propose a quantitatively consistent description of one-particle
transport in the network based on observation that the
transmission of an electron across the wells, from one quantum
wire to the other, happens due to the excitation of oscillatory
modes in the well. This approach permits us to obtain an
approximate formula for the transmission coefficients based on
numerical data for eigenvalues and eigenfunctions of the discrete
spectrum of some intermediate Schr\"{o}dinger operator on the
network in certain range of energy. The  obtained approximate
formula  for the transmission coefficients  permits to reduce the
region of search in the space of parameters in course  of
optimization of the design. We interpret the corresponding
approximate Scattering matrix as a Scattering matrix of the
relevant solvable model which can be used when modelling the
network by a quantum graph. }
\end{abstract}
     \vskip5pt



 \maketitle

\section{Introduction}

Modern  interest to  quantum networks  is inspired  by the
engineering of quantum electronic  devices and  oriented to
manufacturing of networks with prescribed  transport properties.
In simplest case  of a single-particle, single-mode  processes
transport properties can be  described in terms of  Scattering.
Quantum conductance was related to Scattering Processes in
pioneering papers  by Landauer and B\"{u}ttiker, see
\cite{Landauer70,Buttiker85}, and the role of the resonance
scattering in mathematical design of quantum electronic devices
was  clearly understood  by the beginning  of the nineties, see
\cite{Adamyan,BH91,Buttiker93}.Nevertheless practical design of
devices  up to now, see for instance \cite{Compano}, was based on
the formal resonance of energy levels rather than on resonance
properties of the corresponding wave functions. At the same  time
importance of interference in  mathematical design of devices was
noticed in \cite{Exner88,Alamo} and intensely  studied in
\cite{PalmThil92,Exner96,Interf1,Interf2}, see also recent papers
\cite{Safi99,Aver_Xu01,ScattXu02,Kouw2002}.
\par
Experimental technique already permits to observe resonance
effects caused by the shape of the resonance wave functions, see
\cite{B1,B2,Carbon01}. In  our previous papers, see for instance
\cite {Helsinki2,PRB,boston}, we proposed using the shape
resonance  effects as a tool for manipulation of the quantum
current. In particular, we considered the process of resonance
transmission of an electron in the  switch across the quantum well
from one quantum wire to another based on excitation of resonance
modes inside the vertex domain- a phenomenon which was first
mathematically described in \cite{Opening}.
\par
 In actual paper we
apply this basic idea to description of  one-electron transmission
in general quantum networks which can be formed on the surface of
a semiconductor of quantum wells and  quantum wires joining them
or connecting them to  infinity. The depth of the wires and the
wells with respect to the potential on the complement of the
network is assumed large enough to replace the matching boundary
conditions on the boundary of the network by homogeneous Dirichlet
conditions for electrons with energy close to the Fermi level $
E_{_{F}} = \Lambda $ in the wires, see \cite{Madelung}. The
limitations we impose on the width $\delta $ of the wires are
practically {\it not too strict}. They are defined from comparison
of the non-dimensional wave-number of electron in the wires on
Fermi level $\Lambda$ and the non-dimensional inverse spacing
between eigenvalues of an intermediate operator on the resonance
level $\lambda \approx \Lambda$. The numerical analysis of the
circular quantum switch, see \cite{boston}, shows that our
approximation is applicable to the network  constructed of a
circular vertex domain  with diameter $d$ and  quantum wired width
$\delta$ submitted to the condition $d \geq 4 \delta$. Hence our
further assumptions on the width of the wires do not mean actually
that the wires are ``thin'' or the connection of the wires to the
quantum wells is weak, though we  include below, lemma 3.1, also
analysis of this case. We assume that the dynamics of the
electrons in the wires is single-mode and ballistic on large
intervals of the wires compared with the size of geometric details
of the construction (the width of the wire or the size of the
contacts). Our methods in this paper are  based on analysis of the
one-body scattering problem for the relevant Schr\"{o}dinger
equation on the network.
\par
A general Quantum network which is studied in this paper is a
composite domain of a sophisticated form, a sort of a ``fattened
graph", see for instance \cite{KZ01}. It is impossible to obtain
an explicit  solution of the Schr\"{o}dinger equation on this
domain in analytic form. On the other  hand, the direct
computation is not efficient for optimization of the construction
of the network with prescribed transport properties, because the
resonance transmission effect is observed  only at a multiple
point in the  space of geometrical and  physical parameters.
\par
Substitution  of  the network by proper one-dimensional graph with
special boundary  conditions at the  vertices looks like an
attractive program aimed  to  obtaining explicit  formulae for
solutions  of the  Schr\"{o}dinger equation  on the  network, see
\cite{Gerasimenko,Novikov,Schrader,MP00}. Really, analysis  of the
one-dimensional Schr\"{o}dinger equation on the graph , or even on
some hybrid  domain as in \cite{MP01,MPP02} is a comparatively
simple mathematical alternative, but estimation of the error
appearing  from the substitution of the network by a corresponding
graph is difficult. Asymptotic  behavior of  discrete spectrum in
composite  domains  with  shrinking  channels was studied  in
\cite{Jimbo} see  also  references therein.  Simultaneous
shrinking joining channels and vertex  domains of a realistic
compact network was  studied in extended series  of papers on the
edge of last century. In particular in \cite{KZ01,RS01} the
authors develop, based on \cite{Schat96}, a variational technique
for discrete spectrum of Schr\"{o}dinger operator on a ``fattened
graph". They noticed that the (discrete) spectrum of the Laplacian
on a system of {\it finite length} shrinking wave-guides, width
$\varepsilon$, attached to the shrinking vertex domain diameter
$\varepsilon^{\alpha},\, 0< \alpha < 1 $ tends to the spectrum of
the Laplacian  on the corresponding one-dimensional graph but with
different boundary conditions at vertices depending on  speed of
shrinking. In  case of ``small protrusion" $1/2 < \alpha < 1 $ the
spectrum in the corresponding composite domain tends to the
spectrum on a graph with Kirchhoff boundary conditions at
vertices; for the intermediate case $\alpha = 1/2$ the boundary
condition becomes energy-dependent, and  for ``large protrusion",
$0 < \alpha < 1/2 $, the spectrum on a system of wave-guides, when
shrinking, tends to the spectrum on the graph with  zero boundary
conditions at the points of contact so the vertices play the role
of ``black holes" according to \cite{KZ01}.
\par
In distinction from the papers \cite{KZ01,RS01} quoted above, our
theoretical analysis of the  Schr\"{o}dinger  equation on the
``fattened graph'' \cite{P02} was based on a {\it single-mode
one-body scattering problem} in the corresponding  composite
domain with several semi-infinite wires. The aim of actual paper
is: an ``approximate" description of transport properties of the
network. We achieve this aim via derivation of an explicit formula
for  the  Scattering Matrix in terms of the Dirichlet-to-Neumann
map of a  specific {\it intermediate operator} obtained via proper
splitting ( see \cite{Glazman}) of the relevant Schr\"{o}dinger
operator ${\cal L}$ on the  whole network in a  certain interval
of energy. Construction of the system of intermediate operators
$\left\{L_{_{\Lambda}}\right\}$ is defined by the band structure
of the absolutely continuous spectrum $\sigma_{_a}$ of the
Schr\"{o}dinger operator  on  the network the following way.
\par
The multiplicity $n = n _{_{\lambda}}$  of the absolutely
continuous spectrum of the  Schr\"{o}dinger operator on the
network is finite on each finite interval. If the Fermi level
$E_{_{F}} = \Lambda$ is sitting  on the spectral band $\Delta_n =
\left[T_{n},\,\, T_{n+1}\right]$ with the spectral multiplicity
$n= n (\Lambda)$, then the specific splitting ${\cal
L}\longrightarrow {\cal L}_{_{\Lambda}} = {L}_{_{\Lambda}}\oplus
l_{_\Lambda} $ is applied to the Schr\"{o}dinger operator on the
whole network  to  create the  {\it Intermediate operator}
${L}_{_{\Lambda}}$, defined  in the  orthogonal complement  of
``open channels" in the  wires, see section 2. When matching the
outgoing solutions of the corresponding intermediate
Schr\"{o}dinger equation with exponentials in the chopped-off open
channels, we obtain scattered waves of  the  Schr\"{o}dinger
operator on the whole quantum network and  calculate the
corresponding Scattering matrix.  The eigenvalues of the
intermediate operators $L_{_{\Lambda}}$ below the threshold
$T_{n+1}$ generically  give rise to resonances of the
Schr\"{o}dinger operator on the  whole network, see the formulae
(\ref{Scatt}) and (\ref{Sapprox}) below, or become embedded
eigenvalues. In simplest case,when just one quantum well with four
equivalent wires attached is involved, see \cite{boston}, and only
one simple {\it resonance} eigenvalue $\lambda_{_0}^{^{\Lambda}}$
of the Intermediate operator  is sitting on the first spectral
band $\Delta_4$, close to the Fermi-level $ \Lambda$, the
Scattering matrix is presented approximately by the corresponding
resonance factor (\ref{Sapprox1}). This factor contains the pole
of the ${\cal D}{\cal N}$-map at the resonance eigenvalue and the
normal derivatives of the corresponding eigenfunction $\varphi_0$
projected onto  the corresponding four-dimensional entrance
subspace $E^{^{\Lambda}}_{_{+}} = E_{_{+}}$ of the {\it open
channels},- the  linear hull of  the  corresponding cross-section
eigenfunctions  in the  wires:
\[
S (\lambda) = -\frac{{\cal D}{\cal N}^{^\Lambda} + i K_+} {{\cal
D}{\cal N}^{^\Lambda} - i K_+} \approx
\]
\begin{equation}
\label{Sapprox1} \displaystyle - \frac{\frac{P_{+}\frac{ \partial
\varphi_{_0} (x)}{\partial n_x}\rangle \,\,\langle P_{+}\frac{
\partial \varphi_{_0} (s)}{\partial n_s}} {\lambda -
\lambda^{^{\Lambda}}_{_{0}} } + i K_{+} (\lambda^{^{\Lambda}}_0)}
{\frac{P_{+}\frac{ \partial \varphi_{_0} (x)}{\partial n_x}\rangle
\,\,\langle P_{+}\frac{ \partial \varphi_{_0} (s)}{\partial n_s}}
{\lambda - \lambda^{\Lambda}_0}-i K_{+} (\lambda^{^{\Lambda}}_0)}.
\end{equation}
Here $K_{_{+}}$ is the exponent obtained from  oscillatory
solutions in semi-infinite quantum wires and $P_{+}\frac{ \partial
\varphi_{_0} (x)}{\partial n_x}\rangle \,\,\langle P_{+}\frac{
\partial \varphi_{0} (s)}{\partial n_s}$ is the resonance  term
associated with the resonance eigenvalue
$\lambda^{^{\Lambda}}_{_0}$ and the  corresponding  eigenfunction
$\varphi_0$. It can be presented in standard form $\parallel
P_{+}\frac{ \partial \varphi_0 }{\partial n}
\parallel^2 P_0$ with the orthogonal projection  $P_0$ onto the
one-dimensional subspace $E^{^0}_{+}$ in  $E_+$ spanned by
$P_{+}\frac{ \partial \varphi_0 }{\partial n} $.
\par
  The  above  ``one-pole'' approximation  (\ref{Sapprox1})
of the  Scattering matrix  near the resonance eigenvalue
$\lambda^{^{\Lambda}}_{_{0}} $ is obtained via substitution of the
Dirichlet-to-Neumann map of the intermediate operator  by  the
resonance term only, neglecting all  non-resonance terms and the
contribution from the continuous spectrum. In case of the
resonance triadic Quantum Switch, see \cite{boston}, similar
neglecting was possible in a small neighborhood of the resonance
eigenvalue. The size of the neighborhood was defined by the
``natural" small parameter $\approx 1/20 $ which appears from
above mentioned comparison of the inverse spacing of eigenvalues
of the intermediate operator  on the resonance level and the
wave-number of the electron in the wire on the resonance energy.
We  derive also a ``few-pole approximation" of the  Scattering
matrix and interpret it  as  a  Scattering matrix of  some
solvable model of the  network in certain range of  energy.
\par
Actual  paper has the  following plan: after the  formal
description of the quantum  network and introduction of an
intermediate operator in the  next  section 2, we prove in section
3 that the intermediate operator $L_{\Lambda}$ is self-adjoint,
hence possess a resolvent kernel which has usual properties.Based
on these properties  we introduce the corresponding
Dirichlet-to-Neumann map of the  Intermediate operator and connect
it with  the  Dirichlet-to-Neumann map of  the Schr\"{o}dinger
operator on the Quantum well. The following section 4 contains the
calculation of the Scattering matrix and derivation of the
corresponding ``one-pole approximation''. Then in section 5 we
suggest an interpretation of the ``one-pole approximation''of the
Scattering matrix on  Quantum network as a  Scattering matrix of a
solvable model presented as  a  star-shaped graph with  a single
resonance  node. In the last section we discuss briefly some
historic details of the perturbation theory on the continuous
spectrum.
\par
The  aim of our paper is the description of an approach  to the
transport problem on the  network  based  on the  idea  of the
intermediate  operator. This  approach is demonstrated in details.
But the total volume of this paper is not sufficient to discuss
all subtle mathematical details of each technical step. Hence in
couple of places, which are clearly indicated, see for instance
discussion of the  singular spectrum in  Theorem 3.1, we supply
just a sketch of the corresponding proof. We plan to return to
these questions soon in following publications. \vskip1cm
\section{Intermediate Operators}
\noindent Consider  a  network  on a  plane manufactured  of
several non-overlapping quantum wells (vertex domains)
$\Omega_t,\,\,t=1,2,\dots T$ and a few straight finite or
semi-infinite wires $\omega^m,\,\, l= 1,2,\dots M$ of  constant
width  $\delta_m$ and length $d_m \leq \infty$ attached  to the
domains $\Omega_{t_{1}},\,\,\Omega_{t_2},\dots $  such that  the
bottom sections $\gamma^m_{t_1},\,\gamma^m_{t_2}$ of  the  wire
$\omega^m$ are parts of the piece-wise smooth boundaries $\partial
\Omega_{t_1},\,\,
\partial \Omega_{t_2},\dots $
of  the  domains  $\Omega_{t_1},\,\, \Omega_{t_2},\dots $
respectively. Each  wire is attached to one (if  the wire is
semi-infinite ) or two  domains (if it is  finite), so that the
function $m\to t$ or $m \to t_1,t_2$ is  defined although the
inverse function may not be defined, since there may be a few
wires $\omega^m$ connecting two given domains $\Omega_{t_1},\,
\Omega_{t_2}$. In our further construction we use slightly
extended vertex domains with  small ``cut-off's'' $\delta
\omega^{^{m}}$ of the  wires $\omega^{^{m}}$ attached to them:
$\delta \omega^{^{m}} = \left\{ 0< x < \varepsilon^{^m}\right\}$,
to avoid a discussion of the  subtle  question  of Sobolev
smoothness of restrictions of elements from the  domain  of the
operator  onto the bottom sections of  channels  joining  the
inner  angles. Denote by $\gamma^{^m}_{_t}$ the orthogonal bottom
sections of the ``shortened'' wires  with cut-off's removed (from
both ends,if the wire is finite) and attached to the corresponding
vertex domains. We will use further former notations
$\Omega_{_t},\,\omega^{^m}$ for extended  vertex domains and for
the shortened wires respectively. The previously described vertex
domains and  wires will not  be used in the following text. We
consider the spectral problem for the Schr\"{o}dinger operator
\begin{equation}
\label{Schredinger} {\cal L} u = -\bigtriangleup_{\mu} u + V(x)u =
\lambda u
\end{equation}
on the  network $\Omega = \cup_{t}\Omega_t \cup_m \omega^m$ with
zero boundary condition on  $\partial \Omega$. The kinetic term
$-\bigtriangleup_{\mu}$ containing the tensor  $\mu$ of  effective
mass is  defined as
\[
\displaystyle -\bigtriangleup_{\mu} = \left\{
\begin{array}{cc}
\label{mass}
-(2 \mu_t)^{-1}\bigtriangleup,\,&\mbox{if} \,\in \Omega_t\\
-\frac{1}{2\mu^{^{\parallel}}_{_{m}}}\frac{\partial^2}{\partial
x^2} -\frac{1}{2\mu^{^{\bot}}_{_{m}}}\frac{\partial^2}{\partial
y^2} &\mbox{if} \,\in \omega^m
\end{array}
\right.,
\]
where $\mu_t $ plays the role of the  average effective  mass  in
the well $\Omega_t$, $x,y$ are the coordinates along and  across
the wire $\omega^m$ and
$\mu^{^{\bot}}_{_{m}},\,\mu^{^{\parallel}}_{_{m}}$ are  positive
numbers playing  the  roles  of effective masses across and  along
the wire $\omega^m$ respectively. We  impose the  Meixner
condition at the inner corners of the boundary of the domain in
form $D({\cal L})\subset W_2^1 (\Omega)$. We assume that the
potential $V$  is constant on the wires $V \bigg|_{\omega^{^m}} =
V_m$ and is a real bounded measurable function
    $V_t$ on each (extended) vertex  domain  $\Omega_t$. Without
loss of generality we may assume  that the overlapping wires do
not interact, otherwise we  may treat the overlapping as an
additional quantum well. The absolutely-continuous spectrum of the
operator  ${\cal L}$ coincides with the absolutely- continuous
spectrum of the restriction $\sum_{m = m(t)}\oplus l_m$ of the
operator ${\cal L}$ onto the  sum of  all semi-infinite  wires
$\omega^m,\,\, m = m(t)$ with  zero boundary conditions on  the
boundary $\cup_m \partial \omega^{m}$. This  fact  follows  from
the Glazman's splitting  technique, \cite{Glazman}, see a
discussion in the proof of  theorem 3.1 in the  next section. The
spectrum of the  restriction  $l_{m}$ consists of a countable
number of branches $\cup_{m=1}^{\infty}\left[\left. V_{_m} +
\frac{s^2 \pi^2} {2 \mu^{^{\bot}}_{_m}\delta^2_m},\,\infty
\right.\right)$
    which correspond to the  oscillating  modes
  $\displaystyle e^m_s (y)\,\,\mbox{exp} \pm i\,
  \sqrt{2 \mu^{^{\parallel}}_{_m}}\sqrt{\lambda -
\left[ V_{_m} + \frac{s^2 \pi^2} {2
\mu^{^{\bot}}_{_m}\delta^2_m}\right]}\,\,\,x,\,\,\,
    V_{_m} + \frac{s^2 \pi^2}
{2 \mu^{^{\bot}}_{_m}\delta^2_m}\,\,\, < \,\lambda\, $ spanned by
the eigenfunctions of the cross-section $\displaystyle e^m_s(y) =
\sqrt{\frac{2}{\sqrt{2 \mu^{^{\bot}}_{_m}}\delta_m}} \sin \frac{s
\pi y}{\sqrt{2 \mu^{^{\bot}}_{_m}}\delta_m},\,\, s= 1,2,\dots $.
If the Fermi-level $\Lambda$ does not  coincide  with any
threshold, $\Lambda \neq  V_{_m} + \frac{ s^2 \pi^2} {2
\mu^{^{\bot}}_{_m}\delta^2_m} ,\,\, s = 1,2,\dots$, then the
linear hulls of eigenfunctions of the  cross-sections
corresponding to the open channels $  V_{_m} + \frac{s^2 \pi^2} {2
\mu^{^{\bot}}_{_m}\delta^2_m} < {\Lambda} $ are  considered as
entrance subspaces  of ``open (lower) channels'':
    $E^m_{+}(\Lambda) = \bigvee_{s \pi \leq \sqrt{2
\mu^{^{\bot}}_{_m}(\Lambda-V_{_m})}\,\delta_m} e^m_s $. The
entrance subspaces of ``closed (upper) channels'' $E^m_{-}
(\Lambda) = \bigvee_{s \pi > \sqrt{2 \mu^{^{\bot}}_{_m} (\Lambda -
V_{_m})}\, \delta_m} e^m_s = L_2 (\gamma^m) \ominus
E^m_{+}(\Lambda)$, correspond to the  upper thresholds: $\frac{s^2
\pi^2}{2 \mu^{^{\bot}}_{_m}\delta^2_m} + V_{_m} > \Lambda $.
    Denote by  $P^m_{\pm} (\Lambda)$ the  orthogonal projections onto
the  subspaces $E^m_{\pm} (\Lambda) \subset L_2 (\gamma^m_t)$ on
the  bottom sections of  the  finite and/or infinite  wires
$\omega^m$ lying on  the  border of the (extended) vertex domain
$\Omega_t$.
\par
An important feature of the above quantum network is the
``step-wise'' structure of  the continuous  spectrum  of the
operator  ${\cal L}$ with spectral bands of different spectral
multiplicities separated  by  thresholds, see \cite{Kopilevich}.
This property permits us to introduce the corresponding  structure
of intermediate operators.
\par
Note that elements from the domain of the Schr\"{o}dinger operator
${\cal L}$ on the  whole  network $\Omega$ belong locally (outside
a small  neighborhood of  inner corners) to the proper Sobolev
class $ D ({\cal L}) \subset W_{_2}^{^{2}} (\Omega_t,loc)$, see
\cite{KL88}. For given Fermi level $\Lambda > 0$ we define an {\it
intermediate operator},\,\, following \cite{Helsinki2,boston,PRB}
where the suggested construction was introduced for star-shaped
Quantum Switch. The intermediate operator is  defined  by the same
differential expression as the Schr\"{o}dinger operator ${\cal L}$
defined  on $W^{^2}_{_2}$-smooth functions  in the orthogonal sum
of spaces of all square-integrable functions on the components
$\left\{\omega^m,\,\Omega_t\right\}$ of the network with special
boundary  conditions  at the  bottom sections $\gamma^m \in
\partial \Omega_t,\, t= 1,2, \dots $ of the wires $\omega^m$:
\[
P^m_{+}(\Lambda) u_t \bigg|_{\gamma^m_t} = 0,\,\,\,\,\,
P^m_{+}(\Lambda) u^m \bigg|_{\gamma^m_t} = 0
\]
\begin{equation}
\label{boundcond} P^m_{-}(\Lambda)\left[ u^m \bigg|_{\gamma^m_t} -
    u_t \bigg|_{\gamma^m_t}\right]= 0,\,\,\,\,\,
P^m_{-}(\Lambda)\left[
\frac{1}{2\mu_m^{^{\parallel}}}\frac{\partial u^m}{\partial n}
\bigg|_{\gamma^m_t}
        - \frac{1}{2 \mu_{_t}}\frac{\partial u_t }{\partial n}
\bigg|_{\gamma^m_t} \right]= 0.
\end{equation}
 These boundary  conditions are  meaningful
due to Embedding theorems, see \cite{Embedding},  because
functions from both $W_2^2(\Omega_t,loc),\,\, W_2^2 (\omega^m)$
have boundary values in proper sense on  the elements  of the
common boundary $\partial \omega^m \cap \partial \Omega_t$ - on
the bottom sections $\gamma_{_t}^{^m}$ of the (shortened) wires.
We also  assume  that the  Meixner conditions are fulfilled  at
the inner corners  for the component $u_{_t}$  of the element in
the vertex domains. The obtained  split operator is denoted
further in this section by $\hat{\cal L}_{_{\Lambda}}$, and the
whole domain of it in $\cup_{_t} W_2^2(\Omega_t,loc)\oplus
\cup_{_m} W_2^2 (\omega^m)$ is denoted by  $D\left(\hat{\cal
L}_{_{\Lambda}}\right)$. The  Meixner condition can be  presented
as  $D\left(\hat{\cal L}_{_{\Lambda}}\right)\subset W_{_2}^{^1}
(\Omega)$.
\par
  To present the above boundary conditions in a more compact
form we can introduce orthogonal sums of upper (closed) and lower
(open) entrance spaces $\sum_{m}\oplus \left(E^m_t\right)_{_{\pm}}
:= E_{t,\pm}$ on the wires attached to the  domain $\Omega_t$. We
introduce also similar notations  for the  corresponding
projections $P_{t,\pm} {\bf u}^{\omega}$ of the vectors $\sum_{m}
u^m := {\bf u}_t^{\omega}$ obtained via restriction of functions
defined on the wires attached to the given domain $\Omega_t$ onto
the bottom sections $\gamma_t^m \in \partial \Omega_t\cap\partial
\omega^m$. Similarly we can introduce the restrictions of the
functions $u_t$ defined in $\Omega_t$ onto the sum
    $\Gamma_t = \cup_m\gamma_t^m$ of all bottom
sections $\gamma_t^m$ of the wires $\omega^m$ attached to the
domain $\Omega_t$:
\[
{\bf u}_{_{_{\Omega_t}}} = u_t \bigg|_{_{_{\Gamma_t}}}
\]
and the corresponding projections $ \sum_{_{m}} P^{^m}_{_{t,\,
\pm}} = P_{_{t,\, \pm}}$, and  then form an orthogonal sum:
\[
{\bf u}_{_{_{\Omega}}}= \sum_t \oplus u_{_{\Omega_t}},\,\,P_{\pm
}{\bf u}_{_{_{\Omega}}} = \sum_{t,m}\oplus  P^{^m}_{_{t,\pm }}
{\bf u}_{_{\Omega_t}}.
\]
Now the  above boundary  conditions can be  re-written  in a
compact form  as
\begin{equation}
\label{bcond} P_{+}{\bf u}_{_{_{\Omega}}}= P_{+}{\bf u}^{\omega} =
0,\,\, P_{-}\left[{\bf u}_{_{_{\Omega}}}-{\bf u}^{\omega}\right] =
0,\,\, P_{-}\left[\frac{1}{2{\bf
\mu}_{_{\Omega}}}\frac{\partial{\bf u}_{_{_{\Omega}}}}{\partial n}
- \frac{1}{2{\bf \mu}^{^{\parallel}}}\frac{\partial{\bf
u}^{\omega}}{\partial n}\right] = 0.
\end{equation}
Here $\mu_{\Omega},\, \mu^{\parallel}$  are diagonal tensors
composed of  values of the average effective masses in domains
$\Omega_t$ and along the wires $\omega^{m}$ respectively. The
boundary condition \ref{bcond}  may be  interpreted  as a
``chopping-off'' condition  in  all open (lower)  channels
$\frac{s^2 \,\pi^2}{2 \mu^{\bot}_{_{\omega}}\delta_m^2} + V_{_m} <
\Lambda $ and  a ``partial matching  condition'' in all closed
(upper)  channels $\frac{s^2\,\pi^2}{2 \mu^{\bot}_{_{\omega}}
\delta_m^2} + V_{_m}
>  \Lambda $.
    The restrictions $\hat{l}^m (\Lambda)$ of  the split operator
$\hat{\cal L}$ onto the invariant subspaces ${\cal H}^m_+ =
\bigvee_{\frac{s^2\,\pi^2}{2 \mu^{\bot}_{_{\omega}}\delta_m^2} +
V_{_m} < \Lambda} e^m_s
    \times L_2(0,\, d_m)
= E^m_{+} \times L_2(0,\, d_m)$
    which correspond to the (open) channels
    $\frac{s^2\,\pi^2}{2 \mu^{\bot}_{_{\omega}}\delta^2_m} + V_{_m}< \Lambda$,
are self-adjoint operators and admit spectral analysis in explicit
form. We denote by $\hat{l}_{_{\Lambda}}$ the  orthogonal sum of
all  operators $\hat{l}^{^{m}}_{_{s}}(\Lambda)$ in all open
channels. The restriction $\hat{L}_{\Lambda}$ of  $\hat{\cal L}$
onto the complementary invariant subspace is  also  a self-adjoint
operator :
\[
\hat{L}_{\Lambda} = \hat{\cal L}\,\ominus\, \left\{\sum_{\frac{s^2
\,\pi^2}{2 \mu^{\bot}_{_{\omega}}\delta^2_m} + V_{_m} <
\Lambda}\,\,\, {\oplus}\,\,\,\hat{l}^{^{m}}_{_{s}}(\Lambda)
\right\}:= \hat{\cal L}\,\ominus\, \hat{l}_{_{\Lambda}}.
\]
The above chopping-off construction can be applied  either to open
channels  in  all  wires, both finite and infinite, resulting in
the  above operators $\hat{L}_{_{\Lambda}}$ and
$\hat{l}_{_{\Lambda}}$, or applied to the semi-infinite open
channels only, resulting in  corresponding operators
$L_{_{\Lambda}},\,\,\, l_{_{\Lambda}}$ respectively,
    ${\cal L}_{_{\Lambda}} = L_{_{\Lambda}}\,\oplus\, l_{_{\Lambda}}$.
One of the intermediate operators : $\hat{L}_{\Lambda} $ or $
L_{_{\Lambda}} $ can be  more  convenient for calculating the
Scattering matrix, depending on  the  architecture of  the
network.
    In  the  remaining part of this  paper we  choose  $L_{_{\Lambda}}$ as an
    intermediate operator, chopping off  only  the  open channels in
semi-infinite  wires.
\par
Some useful  modification  of the  above  chopping-off
construction is also  convenient: for  given  value  $\Lambda$  of
the  Fermi level  one can choose  the number  $\Lambda_{_1} >
\Lambda$ and  split the operator ${\cal L}$ as
 $L_{_{\Lambda_{_{1}}}}\oplus l_{_{\Lambda_{_{1}}}}$  with   $l_{_{\Lambda_{_{1}}}}$,
containing  several  closed  channels, for  given  Fermi level, in
the finite lower group. The formulae  which  are  derived  for the
Scattering matrix in this  case look  slightly less elegant
because of  excessive number of  matching conditions. But an
essential gain of this modification of the  intermediate  operator
is  the  shift of the lover  border of the continuous  spectrum of
the intermediate operator  beyond  the   level  $\Lambda_{_1}$, at
the minor price of  appearance  of a  finite number  of
eigenvalues the intermediate operator $L_{_{\Lambda_1}}$ below
$\Lambda_{_1}$. This modification of the  construction of the
intermediate operator will  be  used  in course of  the  proof  of
the Theorem 3.1 and further in section 5, see also a relevant
``physical" motivation in  the beginning of section $5$. We
postpone the standard derivation of the corresponding formulae for
the Scattering matrix to the following publication, but
concentrate further mainly  on the contribution to the Scattering
matrix from the discrete spectrum of $L_{_{\Lambda}}. $\vskip1cm
\section{Dirichlet-to-Neumann map}
When studying Scattering on one-dimensional Quantum Networks one
obtains the Scattered waves of the operator ${\cal L}$ on the
network via matching solutions of the split  Schr\"{o}dinger
operator ${\cal L}_{_{\Lambda}}$ with elementary exponentials,
  or, generally, Jost functions, in
the semi-infinite wires at infinity, see for instance
\cite{Carlson98,Gerasimenko,Harmer02,Solomyak}. In case of more
realistic two-dimensional wires  matching of solutions from
neighboring domains on common boundary is also an initial step of
the perturbation procedure.
\par
Now  we focus on the  multi-dimensional techniques of matching
based on a special version of the Dirichlet-to-Neumann map
\cite{SU2,DN01} modified for the intermediate operator on the
Quantum Network. We need, first of all, the   self-adjointness of
the intermediate operator, the corresponding Green function and
the Poisson map. Then, based on them, we suggest an algorithm for
the construction of  the Dirichlet-to-Neumann map of the
intermediate operator which gives eventually the corresponding
Scattering matrix.
\par
Most of facts collected in the following statement are very well
known, but scattered in literature, beginning from \cite{Rellich}.
Lot's of important analytic facts we need, and even in stronger
form, can be found in \cite{Kopilevich,KL88}.

    \begin{theorem}\label{intermed} {\it Consider the  operators
${\cal L}_{_{\Lambda}},\,L_{_{\Lambda}},l_{_{\Lambda}},\,\, {\cal
L}_{_{\Lambda}} = L_{_{\Lambda}} \oplus l_{_{\Lambda}},$ defined
by the differential expression (\ref{Schredinger}), the above
boundary conditions (\ref{bcond}),or just matching conditions on
$\Gamma$, and Meixner condition at the inner corners of the
boundary  of the whole network. The domains of the operators
consist of all properly smooth functions defined on the sum of all
open semi-infinite channels ${\cal H}_{+} (\Lambda) =
\sum_{m}{\cal H}^m_{+}(\Lambda)$  and in the orthogonal complement
${\cal H}_{-} (\Lambda) = L_2 (\Omega)\ominus \sum_{m} {\cal
H}^m_{+} (\Lambda)$ respectively. The absolutely-continuous
spectra $\sigma (l_{_{\Lambda}}),\,\sigma ({\cal L}_{_{\Lambda}})
$ of the operators $l_{_{\lambda}},{\cal L}_{_{\lambda}} $
coincide with the  joining of all lower and upper branches in
semi-infinite wires, respectively:
\[
\sigma (l_{_{\Lambda}}) = \cup_{_{\left\{\frac{s^2 \,\pi^2}{2
\mu^{\bot}_{_m} \delta^2_m} < \Lambda- V_{_m} \right\}}}
\left[\left.\frac{s^2 \,\pi^2}{2 \mu^{\bot}_{_m}\delta^2_m} +
V_{_m},\,\infty \right. \right),
\]
\[
\sigma (L_{_{\Lambda}}) = \cup_{_{\left\{\frac{s^2 \,\pi^2}{2
\mu^{\bot}_{_m} \delta^2_m} > \Lambda- V_{_m} \right\}}}
\left[\left.\frac{s^2 \,\pi^2}{2 \mu^{\bot}_{_m}\delta^2_m} +
V_{_m},\,\infty \right. \right).
\]
The  absolutely-continuous  spectrum  $\sigma({\cal L})$ of the
operator ${\cal L}$ coincides with  the  absolutely-continuous
spectrum  of the  the  orthogonal sum  ${\cal L}_{_{\Lambda}} =
L_{_{\Lambda}} \oplus l_{_{\Lambda}}$ :
\[
\sigma({\cal L}) = \sigma ({\cal L}_{_{\Lambda}}) = \sigma
(l_{_{\Lambda}}) + \sigma (L_{_{\Lambda}})
\]
Besides the  absolutely  continuous  spectrum, the  operators
${\cal L}$ and $L_{_{\Lambda}}$ may have a finite number of
eigenvalues below the threshold of the absolutely-continuous
spectrum and a countable sequence of embedded eigenvalues
accumulating at infinity. The singular continuous spectrum of both
${\cal L},\, {\cal L}_{_{\Lambda}}$ is absent.
  \par
The original  operator ${\cal L}$ on the  whole network is
obtained from the split operator ${\cal L}_{_{\Lambda}} =
L_{_\Lambda}\oplus{l_{_{\Lambda}}}$ via the finite-dimensional
perturbation replacing the first of the boundary conditions
(\ref{bcond}) by  the corresponding partial matching boundary
condition in the lower channels.}
  \par
\end{theorem}
{\it Proof}\,\,\,\, To verify the self-adjointness of the operator
${\cal L}_{_\Lambda}$ it is sufficient to prove that the operator
${\cal L}_{_{\Lambda}}$ is symmetric on the domain consisting of
all properly smooth functions satisfying the above boundary
conditions and it's adjoint is symmetric too. Then the operator
$L_{_{\Lambda}}$ is self-adjoint as  a difference of self-adjoint
operators ${\cal L}\ominus l_{_{\Lambda}}$. \par We introduce the
notation $\cup_{m,s} \gamma^m_s:= \Gamma$ and consider the
decomposition of the  network $\Omega =
\left(\cup_{m}\omega^m\right) \,\cup\,\left(\cup_s \Omega_s\right)
$. Then integrating by parts with functions  continuously
differentiable on each component of the above  decomposition we
may present the  boundary  form  of the Schr\"{o}dinger  operator
$ L$ without boundary conditions in terms  of  jumps
$\displaystyle \left[ {\bf u} \right]= u^{\omega}- u^{\Omega},\,
\,\left[ \frac{\partial {\bf u}}{\partial n} \right] =
\frac{1}{2\mu^{^{\parallel}}_{_m}}\frac{\partial
{u}^{\omega}}{\partial n}- \frac{1}{2\mu_{_t}}\frac{\partial {
u}_{\Omega}}{\partial n}$ of the  functions, and ones of the
normal derivatives on  bottom sections of the  wires and  the
corresponding mean values $\left\{ {\bf u} \right\}\,=
\frac{1}{2}\left\{ u^{\omega} + u_{_{_\Omega}}\,\right\},
    \left\{ \frac{\partial {\bf u}}{\partial n} \right\} =
    \frac{1}{2}\left\{\frac{1}{2\mu^{^{\parallel}}_{_m}}\frac{\partial
{u}^{\omega}}{\partial n} + \frac{1}{2\mu_{_t}}\frac{\partial {
u}_{_{_\Omega}}}{\partial n}\right\}$:
    \[
\langle {\cal L} {\bf u},\, {\bf v} \rangle - \langle u,\,{\cal L}
{\bf v} \rangle  =
\]
\[
\int_{\Gamma} \left(\left[\frac{\partial {\bf u}}{\partial
n}\right]\left\{ \bar{{\bf v}}\right\} - \left\{ {\bf u}\right\}
\left[\frac{\partial \bar{{\bf v}}}{\partial n}\right] \right)
    d\Gamma +
\]
\[
\int_{\Gamma} \left(\left\{\frac{\partial {\bf u}}{\partial
n}\right\}\left[ \bar{{\bf v}}\right] - \left[ {\bf u}\right]
\left\{\frac{\partial \bar{{\bf v}}}{\partial n}\right\} \right)
d\Gamma .
\]
Denote  by  $P^{^{\Lambda}}_+$ the  orthogonal sum of orthogonal
projections onto the  sum $E^{^{\Lambda}}_{_+}$  of the entrance
subspaces of the lower (open) channels in $L_2 (\Gamma)$ and by
$P^{^{\Lambda}}_{_{-}}$ the complementary projection ,
    $I = P^{^{\Lambda}}_+ + P^{^{\Lambda}}_{-}$. Inserting
this decomposition of unity into the integration over $\Gamma =
\cup_{m,t} \gamma^m_t$ we  can see  that  the  above  boundary
form is equal to
\[
\int_{\Gamma} \left(P^{^{\Lambda}}_{+}\left[\frac{\partial {\bf
u}}{\partial n}\right]\,
    P^{^{\Lambda}}_{+}\left\{\bar{{\bf v}}\right\}
-  P^{^{\Lambda}}_{+}\left\{{\bf u}\right\}
P^{^{\Lambda}}_{+}\left[\frac{\partial \bar{{\bf v}}}{\partial
n}\right] \right)
    d\Gamma +
\]
\[
\int_{\Gamma} \left(P^{^{\Lambda}}_{-}\left[\frac{\partial {\bf
u}}{\partial n}\right] \, P^{^{\Lambda}}_{-}\left\{\bar{{\bf
v}}\right\} - P^{^{\Lambda}}_{-} \left\{{\bf u}\right\}
P^{^{\Lambda}}_{-} \left[\frac{\partial \bar{{\bf v}}}{\partial
n}\right] \right)
    d\Gamma+
\]
\[
\int_{\Gamma} \left( P^{^{\Lambda}}_{+}\left\{\frac{\partial {\bf
u}}{\partial n}\right\}\,
     P^{^{\Lambda}}_{+}\left[\bar{{\bf v}}\right]
-  P^{^{\Lambda}}_{+}\left[ {\bf u}\right]
    P^{^{\Lambda}}_{+}\left\{ \frac{\partial \bar{{\bf v}}}{\partial
n}\right\} \right) d\Gamma +
\]
\[
\int_{\Gamma} \left(P^{^{\Lambda}}_{-}\left\{\frac{\partial {\bf
u}}{\partial n}\right\},
    P^{^{\Lambda}}_{-}\left[ \bar{{\bf v}}\right]
- P^{^{\Lambda}}_{-} \left[ {\bf u} \right]\,
    P^{^{\Lambda}}_{-} \left\{\frac{\partial \bar{{\bf v}}}{\partial n}
\right\} \right) d\Gamma.
\]
One can see from (\ref{boundcond}) that the operator $L$ supplied
with the above
    boundary conditions is  symmetric. Vice versa, if  $L^+$ is  the
adjoint operator
    then the boundary form with ${\bf u} \in D(L),\, {\bf v} \in D(L^+)$
vanishes. Then
    $P^{^{\Lambda}}_+ \left\{{\bf v} \right\}= 0,\,P^{^{\Lambda}}_+
\left[{\bf u} \right]= 0$
     follows from  arbitrariness
    of values of the normal derivatives of the function
$P^{^{\Lambda}}_+ {\bf v} $ on both sides
    of $\gamma^m_{t}$. Similarly  another two boundary  conditions
(\ref{boundcond})
    for  elements ${\bf v} \in D(L^+)$ may be  verified based on the
arbitrariness of values
    of $P^{^{\Lambda}}_- \left\{ {\bf u}\right\}$ and
     $P^{^{\Lambda}}_- \left\{ \frac{\partial {\bf u}}{\partial n} \right\}$.
    Hence the  adjoint operator is symmetric and thus  self-adjoint.

The restriction  $l_{_{\Lambda}}$ of the  operator  ${\cal
L}_{\Lambda}$ onto the lower channels is obviously a self-adjoint
operator. We call it the  {\it trivial component} of  ${\cal
L}_{_{\Lambda}}$ Hence the restriction $L_{\Lambda}$ of it  onto
the orthogonal complement of  lower channels- the {\it nontrivial
component} of ${\cal L}_{_{\Lambda}}$ - is a  self-adjoint
operator too. The absolutely continuous spectrum $\sigma
(L_{\Lambda})$ coincides with the absolutely-continuous spectrum
of the restriction of the original operator ${\cal L}$ on  the
upper channels  $\sigma (L_{_{\Lambda}}) = \left[
\lambda_{_{min}}^{^{\Lambda}},\,\infty \right)$,
\[
    \lambda^{^{\Lambda}}_{_{min}} =  \mbox{min}_{_{_{
    \frac{\pi^2 \, l^2}{2 \mu^{\bot}_{_m}\delta^2_m} + V_{_m} > \Lambda}}}\,\,\,\,
\frac{\pi^2 \, l^2}{2 \mu^{\bot}_{_m}\delta^2_m} + V_{_m}.
\]
Here  is  a  sketch of the  proof  of  this elementary statement.
Consider a splitting  of  ${\cal L}$ with higher ``formal Fermi
level"  $\Lambda_{_1} >> \Lambda$. The absolutely - continuous
spectrum of the  operator ${\cal L}$ is  the  same as an
absolutely-continuous spectrum of the  split operator, due to the
finite-dimensionality of the perturbation: it is  a sum
$\sigma(l_{_{\Lambda_{_{1}}}})$ of absolutely continuous lower
branches corresponding to the trivial component and  the
absolutely continuous  spectrum  of $L_{_{\Lambda_{_1}}}$. Hence
it suffice to prove, that  the spectrum of  $L_{_{\Lambda_{_1}}}$
below the threshold $\lambda^{^{\Lambda}}_{_{\min}} $  is
discrete.
 \par
Consider inside the  wires a solution of the  homogeneous equation
$L_{_{\Lambda_{_1}}}u = \lambda u $ with $\lambda < \Lambda
<\Lambda_{_{1}}$. This equation admits separation
 of  variables inside the  wires
\[
-\frac{1}{2 \mu^{^{\parallel}}}\frac{\partial^{^{2}} u}{\partial
x^{^2}} -
 \frac{1}{2 \mu^{^{\bot}}}\frac{\partial^{^{2}} u}{\partial y{^2}}
 +  V_{_{m}}  u  = \lambda u,
\]
\[
 u =
 \sum_{_{\frac{1}{2 \mu^{^{\bot}}} \,\,
 \frac{\pi^{^2}l^{^2}}{\delta^{^2}} + V_{_{m}}\geq \Lambda_{_1} }}
 c_{_l}  \sin \frac{\pi l}{\delta\,\sqrt{2 \mu^{^{\bot}}}} e^{^{-
 \sqrt{2 \mu^{^{\parallel}}}
 \sqrt{\frac{\pi^{^2} l^{^2}}{\delta^{^2}\,{2 \mu^{^{\bot}}}} + V_{_{m}} -
 \lambda}\,\,\, x}}.
\]
Then each locally  smooth  solution of  this  equation  is
exponentially
 decreasing  in the  wires $|u|(x,y) \leq C_{_{u}} e^{^{-\sqrt{2 \mu^{^{\parallel}}}
 |\Lambda_{_{1}} - \Lambda|^{^{1/2}}\,\, x}}$. In particular,
 the  Green-function of the  operator  $L_{_{\Lambda}}$ is
 square-integrable,hence defines a bounded operator,
  if  exists for  given  $\lambda < \Lambda_{_1}$. Hence  the
 spectrum  of  the  operator $L_{_{\Lambda_{_1}}}$  does  not have a
 purely continuous  component  on the  interval  $(-\infty,\Lambda_{_1})$.
 Assume  that there  exist  an infinite  sequence of  eigenvalues
 $\left\{ \lambda_{_{r}} \right\}$ which is  convergent to  $\Lambda_{_{2}} \leq
 \Lambda$. Without loss of generality one may assume that
 $\lambda_{_{r}} \leq \frac{\Lambda + \Lambda_{_1}}{2}$ for each $r$.
 The (normalized) eigenfunctions are continuous due  to  embedding  theorems,
\[
\sup  |\varphi_{_r}| \leq   C  \Lambda,
\]
with some dimensional constant  $C$, hence they  are exponentially
decreasing in the wires:
\[
|\varphi_{_r}|(x,y) \leq  C\, \Lambda \, e^{^{- \sqrt{2
\mu^{^{\parallel}}} \sqrt{ \frac{\Lambda_{_{1}} - \Lambda }{2}
}x}},
\]
due  to  maximum principle, and  admit an uniform estimate of
``the tail" in the wires $\int_{_M}^{^{\infty}}
|\varphi_{_r}|^{^{2}} (x,y) dm \leq C^{^2} \,\frac{\Lambda^{^2}}
{\sqrt{ \mu^{^{\parallel}}\,\,(\Lambda_{_{1}} - \Lambda)}} $. Then
any linear combination $\varphi = \sum_{_r} c_{_r} \varphi_{_r}$
of the eigenfunctions at  least admits  the estimate
\[
|\varphi|^{^2}_{_{Lip 2/5}} < C^{^2} |L \varphi|^{^2}_{_{L_{_2}}}
= C^{^2} \sum_{_r} c^{^2}_{_r}|L
\varphi_{_{r}}|_{_{L_{_{2}}}}^{^2} \leq C^{^2}\,
\left(\frac{\Lambda_{_1} +
\Lambda}{2}\right)^{^2}|\varphi|^{^2}_{_{L_{_{2}}}}
\]
due  to  the  Parseval identity, and  the  estimation of the  tail
\[
\int_{_M}^{^{\infty}} |\varphi|^{^{2}} (x,y) \,\,dx\, dy \leq
C^{^2} \,\frac{\Lambda^{^2}}
{\sqrt{\mu^{^{\parallel}}\,\,(\Lambda_{_{1}} - \Lambda)}}.
\]
Now compactness  of the  unit ball  in  the  subspace of all
linear  combinations  of eigenfunctions  $\varphi_{_r}$ follows
 from  embedding theorems  on  the  compact  part $\Omega_{_M }=
\Omega\cup \left\{ x^{^2} + y^{^2} \leq M\right\}$ of the network
and the uniform estimation  of the  tails. The  compactness
implies the  finiteness of the  number of the eigenvalues of
$L_{_{\Lambda_{_1}}}$ below $\Lambda$.
 \par
The  absolutely-continuous  spectrum  of the  operator
$l_{_{\Lambda}}$ has  constant  multiplicity dim$
E_{_+}^{^{\Lambda}}$  on semi-infinite interval
  $ (\lambda_{_{max}}^{^{\Lambda}},\,\infty)$,
where
\[
  \lambda_{_ {max}}^{^{\Lambda}} = \mbox{max}_{_{_{
    \frac{\pi^2 \, l^2}{2 \mu^{\bot}_{_m}\delta^2_m} + V_{_m} < \Lambda}}}\,\,\,\,
\frac{\pi^2 \, l^2}{2 \mu^{\bot}_{_m}\delta^2_m} + V_{_m}.
\]
On the  spectral band  $\Delta_{_{\Lambda}}= \left( \lambda_{_
{max}}^{^{\Lambda}} ,\, \lambda_{_ {min}}^{^{\Lambda}} \right)$
the spectral multiplicity of the  absolutely
  continuous  spectra  of both operators  ${\cal L}_{\Lambda},\, {\cal
L}$ is equal to dim$ E_{_+}^{^{\Lambda}}$. In our case embedded
eigenvalues are possible, but   only  a finite  number of them on
any finite  sub-interval of the continuous  spectrum, see for
instance  \cite{Eastham, Parnovski}.
\par
Singular spectrum of the Schr\"{o}dinger operator with  compactly
supported  potential on a finite joining of standard domains, is,
generically, absent, see the  discussion in \cite{simon}, but the
proof  of  this fact in special cases needs an accurate
investigation. In our special case absence of the singular
continuous spectrum may be derived from  the fact that the partial
boundary conditions imposed on the bottom sections of
semi-infinite wires define the finite-dimensional perturbation of
the operator ${\cal L}$. Recently  absence of  singular spectrum
in wave-guiding systems was  discussed in \cite{Krej6_03,Krej7_03}
by  the method which may be applied in our  situation too.
Neverteless we supply below  a sketch of possible proof based on
classical ideas.
\par
Basic definition of $H$-smoothness in \cite{simon}, volume 4,
XXXIII.7 admits a local formulation for the operator
$A,\,\,R_{_{\lambda}} {\cal H} \subset D(A)$ in Hilbert space
${\cal H}$  with respect to the operator $H,\, R_{_{\lambda}} =
\left( H -\lambda I\right) $, see \cite{Yafaev}. Conditions of
local smoothness are actually derived in the theorem XXXIII.25 of
\cite{simon}. They imply the corresponding local conditions of
absence of the singular spectrum of $H + A$ on the interval
$(a,b)$, if we know that operator  $H$ does  not
  have singular spectrum on the interval
and succeed to verify the local smoothness of the perturbation
$A$. The deficiency elements used in course of the construction of
the split operator for real $\lambda$ below the threshold
$\lambda^{^{\Lambda}}_{_{min}}$ of the  absolutely continuous
spectrum of ${\cal L}$ satisfy  the Helmholtz equation in the
wires and hence decrease exponentially at infinity as
$\exp\left(-[\lambda^{^{\Lambda}}_{_{min}} -
\lambda]^{^{1/2}}x\right)$. Hence the finite-dimensional
perturbation of ${\cal L}_{_{\Lambda}} = { L}_{_{\Lambda}}\oplus
{l}_{_{\Lambda}} \to {\cal L}$ is relatively smooth below the
threshold  $\lambda^{^{\Lambda}}_{_{min}}$ , which implies absence
of the singular spectrum of $H$.
\par
The  last  statement of the theorem is  obvious, see the
corresponding  calculations below  in  course  of derivation the
formula (\ref{Smatrix}) for  the  Scattering matrix in terms of
the DN-map of the  intermediate  operator. This  accomplishes the
proof of the  theorem. \vskip0.5cm \noindent Note that
accumulation of embedded eigenvalues of the operator ${\cal L}$ to
a  finite point is discussed in \cite{Edward}, where also an
example of the Schr\"{o}diner operator with exponentially
decreasing electric field and accumulation of eigenvalues to the
threshold is suggested. Note that presence of slowly-decreasing
potentials  in the  wires implies, generally, an accumulation of
the  embedded eigenvalues to  thresholds. This general fact was
discovered recently in \cite{KNP}. In our  case accumulation of
eigenvalues on any finite  interval of the spectral parameter is
impossible because potentials on the semi-infinite wires are
constant ( the electric field is compactly supported).
\par
Assume  that the  eigenfunctions of the   absolutely-continuous
spectrum of the  operator  $L_{\Lambda}$ and  it's Green function
$G_{\Lambda} (x,s)$ are  already  constructed:
\[
-\bigtriangleup_{\mu} G + V G = \lambda G + \delta(x-s).
\]
Then we can construct the Dirichlet-to-Neumann map (DN-map) of the
operator $L_{\Lambda}$. The  standard  DN-map  is  described in
\cite{SU2,DN01}. The  modified  DN-map of  the  operator
$L_{\Lambda}$ can be  obtained via {\it projection} onto the lower
channels of  the  boundary  current $\frac{1}{2\mu_{_{\Omega}}}
P^{^{\Lambda}}_{+}\frac{\partial u}{\partial n}\bigg|_{_{\Gamma}}$
of the solution of the homogeneous equation $l u - \lambda u = 0$
with the boundary condition $u\bigg|_{\Gamma} = u_{\Gamma}\in
E_{+}^{^{\Lambda}}$, for regular values of the spectral variable
$\lambda$. The solution $u$ tends to  zero at  infinity if $\Im
\lambda\neq 0$. It may be  constructed using the resolvent kernel
$G_{_{\Lambda}} (x,x',\lambda)$ of the  operator $L_{_{\Lambda}}$.
The corresponding  Poisson map  is  given by the  formula:
\[
{\cal P}^{^{\Lambda}} u_{\gamma}(x) = -\int_{\Gamma}
\frac{1}{2\mu_{_{\Omega}}} \frac{\partial
G_{\Lambda}(x,s,\lambda)}{\partial n_s}u_{_{\Gamma}}(s)d s.
\]
Then  the corresponding DN-map  of the  operator $L_{\Lambda}$ is
\[
\frac{1}{2\mu_{_{\Omega}}} \frac{\partial {\cal P}^{^{\Lambda}}
u_{\gamma}}{\partial n} = - \int_{\Gamma}
\frac{1}{(2\mu_{_{\Omega}})^2} \frac{\partial^2
G_{\Lambda}(x,s,\lambda)}{\partial n_x\,\,\partial
n}u_{_{\Gamma}}(s)d s := {\cal D}{\cal N}^{^\Lambda} u_{\Gamma}.
\]
The  normal  current  $\frac{\partial {\cal P}^{^{\Lambda}}
u_{\gamma}}{\partial n}$ belongs to $W_{2}^{1/2 - \varepsilon},\,
\varepsilon > 0$, hence the formal  integral operator
$\int_{\Gamma} \frac{\partial^2 G_{\Lambda}(x,s,\lambda)}{\partial
n_x\,\,\partial n}u_{\Gamma}(s)d s $ is  unbounded
 in $W_{2}^{3/2-\varepsilon}(\Gamma)$.
However, since $E^{^{\Lambda}}_{+}\subset W_{2}^{3/2-
\varepsilon}(\Gamma)$ the projection of the normal current onto
$E^{^{\Lambda}}_{_{+}}$ is bounded from $W_{2}^{3/2 -
\varepsilon}(\Gamma)$ to $W_{2}^{3/2 - \varepsilon}(\Gamma)$,
hence
\[
{\cal D}{\cal N}^{^{\Lambda}} u_{_{\Gamma}} = - P^{^{\Lambda}}_{+}
\int_{\Gamma} \frac{\partial^2 G_{\Lambda}(x,s,\lambda)}{\partial
n_x\,\,\partial n}u_{\Gamma}(s)d s \in
    W_{2}^{3/2-\varepsilon}(\Gamma),
\]
\[
{\cal D}{\cal N}^{^{\Lambda}} : W_{2}^{3/2-\varepsilon}(\Gamma)
\longrightarrow W_{2}^{3/2-\varepsilon}(\Gamma).
\]
According to the  previous  theorem 3.1 there are only a finite
number of eigenvalues $\lambda_r^{\Lambda}$ of the operator
$L_{\Lambda}$ in any compact domain of the complex plane
$\left\{\lambda \right\}$. Then  the  following spectral
representation is  valid  for  the  corresponding  DN-map:
\[
{\cal D}{\cal N}^{^{\Lambda}}(\lambda) =  \sum_r
\frac{1}{(2\mu_{_{\Omega}})^{^{2}}} \frac{P^{^{\Lambda}}_{+}\frac{
\partial \varphi_r (x)}{\partial n_x}\rangle \,\,\langle
P^{^{\Lambda}}_{+}\frac{ \partial \varphi_r (s)}{\partial n_r}}
{\lambda - \lambda^{^{\Lambda}_r}} +
\]
\begin{equation}
\label{DNL} \sum_{V_{m} + \frac{\pi^2 \,l^2}{2 \mu^{\bot}_m
\delta^2_m}> \Lambda} \int_{\frac{\pi^2
\,l^2}{\delta^2_m}}^{\infty} \frac{1}{(2\mu_{_{\Omega}})^{^{2}}}
\frac{P^{^{\Lambda}}_{+} \frac{\partial \varphi_{\rho,l}
(x)}{\partial n_x}\rangle \,\,\langle P^{^{\Lambda}}_{+}
\frac{\partial \varphi_{\rho,l} (s)}{\partial n_s}} {\lambda -
\rho} d\rho,
\end{equation}
where the  normal  derivatives  are calculated  on $\Gamma$ and
the integration is extended over all branches of the
absolutely-continuous spectrum of the operator $L_{\Lambda}$
emerging  from the  upper  thresholds
    $\frac{\pi^2 \,l^2}{2 \mu_{_m}\delta^2_m} + V_{_m} > \Lambda$. The
tensor $\mu _{_{\Omega}}$ of  the  effective mass is diagonal, as
defined in the previous section.  Based on above  analysis one can
prove that $P^{^{\Lambda}}_{_{+}}\frac{\partial \varphi}{\partial
n}\bigg|_{_{\Gamma}} \in W_{_2}^{^{3/2-\varepsilon}}$ and  the
integral and the  series (\ref{DNL}) are convergent in proper
sense  due to the presence of the finite-dimensional projection
$P^{^{\Lambda}}_{_{+}}$ onto the entrance subspace
$E^{^{\Lambda}}_+\subset W_2^{3/2 - \varepsilon}$, see a  similar
reasoning in  \cite{P02}. The whole finite-dimensional
matrix-function ${\cal D}{\cal N}^{^{\Lambda}}(\lambda)$ is
bounded in $W_2^{3/2 - \varepsilon}$ with poles  of first order at
the eigenvalues $\lambda^{^r}_{_s}$ of the operator $L_{_\Lambda}$
and cuts along the upper branches $\left[\frac{\pi^2 \,l^2}{2\,
\mu^{^\bot}_{_m} \delta^{^2}_{_{m}}} + V_{_m}\right.,\left.\infty
\right) $ of the absolutely-continuous spectrum.
\par
Practical  calculation of  the  DN-map  ${\cal D}{\cal
N}^{^{\Lambda}}(\lambda)$ by  the  above  formula (\ref{DNL})
requires  knowing of eigenvalues and  eigenfunctions of discrete
and absolutely-continuous spectrum of the intermediate operator
$L_{\Lambda}$.
\par
 Another  expression for  the  DN-map  may be also
useful. This expression for ${\cal D}{\cal N}^{^{\Lambda}}$ is
obtained in terms of matrix elements of  the DN-map   ${\cal
D}{\cal N}$ of the compact part $\Omega_0$ of  the network with
all semi-infinite wires having zero boundary condition on their
bottom sections.
\par
The  spectrum of the  Schr\"{o}dinger operator $L_0$ on the
``extended" compact part  $\Omega_0$  of the network with  zero
boundary conditions on the bottom sections of the  semi-infinite
channels is discrete. The DN-map of $L_0$ is an operator with  the
generalized kernel
\[
{\cal D}{\cal N} (x,s,\lambda) = -
\frac{1}{(2\mu_{_{\Omega}})^{^2}}
    \frac{\partial^2 G_{_{0}}(x,s,\lambda)}{\partial n_x\,\,\partial n_s}
\]
\[
{\cal D}{\cal N}(\lambda) : W_2^{3/2 - \varepsilon} (\partial
\Omega_0)\longrightarrow W_2^{1/2 - \varepsilon}(\partial
\Omega_0).
\]
We  consider the  restricted  ${\cal D}{\cal N}$-map framed  by
projections on $E = L_{_{2}} (\Gamma)$:
\[
P_{_{E}}{\cal D}{\cal N}(\lambda)P_{_{E}}.
\]
Denoting by $\phi_r (x)$ the  orthogonal projections  onto $E$ of
the boundary  currents $\frac{ \partial \varphi_r (x)}{\partial
n_x}$ of eigenfunctions  of  the  operator  $L_0$ we obtain the
spectral representation of the framed DN-map of the operator $L_0$
by the  formal series
\[
P_{_{E}}{\cal D}{\cal N}(\lambda) P_{_{E}}=  \sum_r
\frac{1}{(2\mu_{_{\Omega}})^{^2}} \frac{\frac{ P_{_{E}}\partial
\varphi_r (x)}{\partial n_x}\rangle \,\,\langle P_{_{E}}\frac{
\partial \varphi_r (s)}{\partial n_r}} {\lambda - \lambda^{0}_r}
: =
    \sum_r \frac{1}{(2\mu_{_{\Omega}})^{^2}}
\frac{ \phi_r (x)\rangle \,\,\langle \phi_r (s)} {\lambda -
\lambda^{0}_r},
\]
which is convergent  in  weak sense on smooth elements  from
$E_{_{+}}$.
\par
 To derive the formula connecting
the DN-maps of the operator $L_{_{0}}$ and one of the intermediate
operator we present the framed DN-map of $L_0$ as a matrix with
respect to the orthogonal decomposition $L_2 (\Gamma) =
E^{^{\Lambda}}_+ \oplus E^{^{\Lambda}}_{-}$ into entrance
subspaces of the  open and closed channels. Based  on the
observation $E_{+}\in W_2^{3/2-\varepsilon}$, we see that  the
matrix elements ${\cal D}{\cal N}_{_{+-}}(\lambda)$ are operators
mapping $E_{\pm}\dots $ into $E_{\pm}\dots$ respectively (for
regular $\lambda$)
\begin{equation}
\label{DN_0}
 P_{_{E}}{\cal D}{\cal
N}P_{_{E}} = \left(
\begin{array}{cc}
P^{^{\Lambda}}_+ \left({\cal D}{\cal N}\right) P^{^{\Lambda}}_{+}&
P^{^{\Lambda}}_+
\left({\cal D}{\cal N}\right) P^{^{\Lambda}}_{-}\\
P^{^{\Lambda}}_- \left({\cal D}{\cal N}\right) P^{^{\Lambda}}_{+}&
P^{^{\Lambda}}_- \left({\cal D}{\cal N}\right) P^{^{\Lambda}}_{-}
\end{array}
\right): = \left(
\begin{array}{cc}
{\cal D}{\cal N}_{_{++}}& {\cal D}{\cal N}_{_{+-}}\\
{\cal D}{\cal N}_{_{-+}}& {\cal D}{\cal N}_{_{--}}
\end{array}
\right).
\end{equation}
Consider the basis $\left\{ e^m_l \right\}$  of all entrance
vectors of closed channels, $\frac{\pi^2 l^2}{2
\mu^{\bot}_{_m}\delta_m^2} + V_{_m}> \Lambda $, and  introduce the
diagonal matrix  $K_{_{-}}^{^{\Lambda}}$ with elements $k^m_l = -
\sqrt{2 \mu^{^{\parallel}}}_{_m}\,\sqrt{\frac{\pi^2 l^2} {2
\mu^{\bot}_{_m}\delta_m^2} + V_{_m} - \lambda},\,\, $ on the
entrance  subspaces $E_{_l,-}^{^{m}}$ of closed  channels
$\frac{\pi^2 l^2}{2 \mu^{\bot}_{_m}\delta_m^2} + V_{_m}> \Lambda
$. The corresponding operator $K_{_{-}}^{^{\Lambda}} (\lambda): =
K_{_{-}} (\lambda)$ has a  bounded inverse $\left(
K_{_{-}}\right)^{^{-1}} : W_2^{1/2-\varepsilon} \to
W_2^{3/2-\varepsilon},\,\,\varepsilon \geq 0$ for  $\lambda $
below the  minimal upper  threshold
    ${\lambda}^{^{\Lambda}}_{_{min}}$.

Let ${\bf u}= \left\{ u_t,\,u^{m}\right\}$ be a  solution of  the
Schr\"{o}dinger  equation  on the  network with the boundary data
$u_{_{\Gamma}} \in E_{_{+}}$ on the sum $\Gamma$ of bottom
sections of the semi-infinite open  channels and matching boundary
conditions in the upper  channels
    \[
P^{^{\Lambda}}_{+}{\bf u}_{\Omega}  = u_{_{\Gamma}},\,\,
P^{^{\Lambda}}_{-}\left[{\bf u}_{\Omega}-{\bf u}_{\omega}\right] =
0,\,\, P^{^{\Lambda}}_{-}\left[\frac{1}{2
\mu_{_{\Omega}}}\,\frac{\partial{\bf u}_{\Omega}}{\partial n} -
\frac{1}{2 \mu^{^{\parallel}}_{_{\omega}}}\frac{\partial{\bf
u}^{\omega}}{\partial n}\right] = 0
\]
and standard  matching boundary  conditions on  all  bottom
sections of the finite  wires. Denote by $u_{-}$ the  projection
of  the  solution  onto  the entrance  subspace  $E_-$ in  the
upper  channels in the semi-infinite  wires.
\par
We calculated above  the  DN-map of the  operator $L_{\Lambda}$ as
a projection
    onto $E_+$  of the boundary  current
\[
{\cal D}{\cal N}^{^{\Lambda}} u_{\Gamma} =
\frac{1}{2\mu_{_{\Omega}}} P_{+} \frac{\partial {\bf u}}{\partial
n}
\]
of  the  outgoing \cite{Lax} solution  $u$ of  the  homogeneous
equation ${\cal L}_{_{\Lambda}} u = \lambda u$ with the  boundary
condition $u \big|_{\Gamma} = u_{_{\Gamma}} \in E_{_{+}}$ and the
matching conditions in upper  channels:
    \[
P_{-}\left[{\bf u}_{\Omega}-{\bf u}_{\omega}\right] = 0,\,\,
P_{-}\left[\frac{1}{2\mu_{_{\Omega}}} \, \frac{\partial{\bf
u}_{\Omega}}{\partial n} - \frac{1}{2
\mu^{^{\parallel}}_{_{\omega}}}\frac{\partial{\bf
u}^{\omega}}{\partial n}\right] = 0.
\]
Due to (\ref{DN_0}) his  gives the  following  system of
equations:
\[
P_+ \left({\cal D}{\cal N}\right) P_{+} u_{\Gamma} + P_+
\left({\cal D}{\cal N}\right) P_{+}  u_{-}=
    {\cal D}{\cal N}^{^{\Lambda}} u_{\Gamma},
\]
\[
P_- \left({\cal D}{\cal N}\right) P_{+} u_{\Gamma} + P_-
\left({\cal D}{\cal N}\right) P_{-} u_{-} = K_{-} u_{-}.
\]
Eliminating  $u_{-}$ we  obtain the  following  statement:
\begin{theorem}{\it The DN-map  ${\cal D}{\cal N}^{^{\Lambda}}$  of
the intermediate operator $L_{\Lambda}$ is  connected  with  the
DN-map $\left({\cal D}{\cal N}\right)$ of the operator  $L_0$ on
the compact part $\Omega_0$ of  the  network $\Omega$ by  the
formula
\begin{equation}
\label{DNr} {\cal D}{\cal N}^{^{\Lambda}} = P^{^{\Lambda}}_+
\left({\cal D}{\cal N}\right) P^{^{\Lambda}}_+  -
    P^{^{\Lambda}}_+ \left({\cal D}{\cal N}\right) P^{^{\Lambda}}_{_{-}}
\frac{I}{P^{^{\Lambda}}_- \left({\cal D}{\cal N}\right)
P^{^{\Lambda}}_- - K_{-}}
    P^{^{\Lambda}}_- \left( {\cal D}{\cal N} \right) P^{^{\Lambda}}_+ .
\end{equation}
}
\end{theorem}
The obtained formula for ${\cal D}{\cal N}^{^{\Lambda}}$ can be
convenient when considering  an independent shrinking
\cite{Kuch02} of vertex domains and  the quantum wires of the
network, if the local geometry of the network permits it. The
DN-map of the Schr\"{o}dinger operator on the system of channels
is contained in (\ref{DNr})  in explicit form, i.e. as
 diagonal matrices of square roots: the negative on the real axis
matrix $K_{_{-}} = $ diag
$\left\{-\sqrt{2\mu^{\parallel}_{_m}}\,\,\sqrt{\frac{\pi^2 l^2} {2
\mu^{\bot}_{_m}\delta_m^2} + V_{_m} - \lambda}\right\}$ and purely
imaginary matrix with a positive imaginary part (on real axis)
\,\,\,$i K_{_{+}} = $\,\,\,i\, diag $ \left\{\sqrt{
2\mu^{\parallel}_{_m}}\,\,\sqrt{\lambda - \frac{\pi^2 l^2} {2
\mu^{\bot}_{_m}\delta_m^2} - V_{_m}} \right\}$. Matrix elements of
the DN-map of the  Schr\"{o}dinger operator $L_{{0}}$ in the
domain $\Omega_0$ are rational  functions
    of  the  spectral parameter with  singularities  at  the
eigenvalues of  the  operator $L_0$.
    \par
{\bf Remark} Note  that the  denominator $\left({\cal D}{\cal
N}\right)_{_{-}} - K_{-} := {\bf D}(\lambda) $ in the above
formula (\ref{DNr}) can be conveniently  transformed  due to  the
invertibility of  $K_{-} = - |K_{-}|$ on the  Fermi-level:
\[
{\bf D}(\lambda) = P^{^{\Lambda}}_- \left({\cal D}{\cal N}\right)
P^{^{\Lambda}}_- - K_{-} = |K_{-}|^{^{1/2}}\left[ I +
|K_{-}|^{^{-1/2}}\left({\cal D}{\cal
N}\right)_{_{-}}|K_{-}|^{^{-1/2}}\right]|K_{-}|^{^{1/2}},
\]
where  the operator in the  bracket is is  acting in
$W^{3/4-\varepsilon}_2$ and  may  be extended  onto $L_2$ by
continuity.
\par
 Our  nearest aim
is  investigation of its  structure and  establishing conditions
of invertibility of the  denominator  ${\bf D}(\lambda)$.
\par
We  can present the operator ${\cal D}{\cal N}$ via Hilbert
identity, see \cite{DN01},  with  the corresponding Poisson
kernels $P_{_{H}}$ for large positive $H$:
\begin{equation}
\label{Hilbert} {\cal D}{\cal N} (\lambda) = {\cal D}{\cal N} (-H)
- (\lambda + H) {\cal P}^{^+}_{_{H}}{\cal P}_{_{H}} - (\lambda +
H)^2 {\cal P}^{^+}_{_{H}} \left(L_{_0} - \lambda I
\right)^{^{-1}}{\cal P}_{_{H}}.
\end{equation}
The  first  summand  acts as an operator from
$W^{3/2-\varepsilon}_2$ to $W^{1/2-\varepsilon}_2$. Hence,
substituting (\ref{Hilbert}) into the preceding formula for the
denominator we notice, due to  $E_+ \subset W^{3/2 -
\varepsilon}_2$, and invertibility of $K_{_{-}}:
W^{1/2-\varepsilon}_2\to W^{3/2 - \varepsilon}_2$, that the
operator $|K_{-}|^{^{-1/2}}\left({\cal D}{\cal
N}\right)_{_{-}}|K_{-}|^{^{-1/2}}$  is  presented  as
\[
|K_{-}|^{^{-1/2}} P^{^{\Lambda}}_{_-} {\cal D}{\cal N}
(-H)P^{^{\Lambda}}_{_-}|K_{-}|^{^{-1/2}}-
\]
\[
- (\lambda + H)|K_{-}|^{^{-1/2}} P^{^{\Lambda}}_{_-}{\cal
P}^{^+}_{_{-H}}{\cal
P}_{_{-H}}P^{^{\Lambda}}_{_-}|K_{-}|^{^{-1/2}} -
\]
\begin{equation}
\label{denom} (\lambda + H)^2
|K_{-}|^{^{-1/2}}P^{^{\Lambda}}_{_-}{\cal P}^{^+}_{_{H}}
\left(L_{_0} - \lambda I \right)^{^{-1}}{\cal
P}_{_{H}}P^{^{\Lambda}}_{_-}|K_{-}|^{^{-1/2}}.
\end{equation}
The  first term  in this  expression is  a  bounded  operator in
$W^{3/2-0}_2$, and may be  extended as a  bounded operator onto
$E= L_{2}(\Gamma) $. Others terms are compact in $E$. Moreover,
one can prove, following similar reasoning in \cite{P02}, that the
operator $|K_{-}|^{^{-1/2}}P_{_-}P^{^+}_{_{H}} \left(L_{_0} -
\lambda I \right)^{^{-1}}P_{_{H}}P_{_-}|K_{-}|^{^{-1/2}}$ is
trace-class operator in $E= L_{2}(\Gamma) $.
\par
 We will use the decomposition of the last term
into the series of polar summands, and  separate the resonance
term $\frac{\phi_0\rangle \langle \phi_0} {\lambda - \lambda_0}$,
with $\phi_0 = P_{_{-}}\frac{\partial \varphi_{_0}}{\partial n}$,
obtained as a projection of the boundary  current of the resonance
eigenfunction $\varphi_{_{0}}$ onto the entrance subspace
$E^{^{\Lambda}}_{_{-}}$ of the closed  channels. Then, omitting
the factors $|K|^{^{-1/2}}_{_{-}}$ we may arrange the summands
according to the decomposition
    \[
{\cal D}{\cal N} =\frac{1}{(2\mu_{_{\Omega}})^2}
\frac{\phi_0\rangle \langle \phi_0} {{\lambda} - {\lambda^0_0}} +
\sum_{s\neq 0}\frac{1}{(2\mu_{_{\Omega}})^2} \frac{\phi_s\rangle
\langle \phi_s} {{\lambda} - {\lambda^0_s}}
:=\frac{1}{(2\mu_{_{\Omega}})^2}
    \frac{\phi_0\rangle \langle \phi_0}
{\lambda - \lambda^0_0} + {\cal K}_{_{0}},
\]
where  ${\cal K}_{_{0}}$ is  the  contribution  to ${\cal D}{\cal
N}$ from the non-resonance eigenvalues  $\lambda^0_s \neq
\lambda^0_0$. The norm of the contribution ${\cal K}_{_{0}}$, as
an operator from $W_2^{3/2- \epsilon}$  to  $W_2^{1/2- \epsilon}$
(for  each $\epsilon >0$), may be  estimated  by  the  spacing
$\rho (\lambda_0)$ at the resonance level $\lambda_0 \approx
\Lambda$. We  explore  the  invertibility of the  denominator
${\bf D}(\lambda)$ in  two  special cases : for  shrinking
networks and  for  thin networks, see  comments  below.
\par
\begin{lemma}{\it If  the  wires are shrinking as  $y \to
\frac{y'}{\varepsilon^{^{\omega}}}$ and the  quantum wells  are
shrinking as $x \to \frac{x'}{\varepsilon_{_{\Omega}}}$, but the
Fermi level is kept constant during the shrinking, then the norm
of the operator  $K^{^{-1}}_{_{-}}\,\,{\cal K}$ on functions of
the variables $x',\, y'$ is  estimated in  $L_2 (\Gamma)$ as
\begin{equation}
\label{shrink}
\parallel K^{^{-1}}_{_{-}}\,\,{\cal K}_{_{0}} \parallel \leq
    C_0 \left(\frac{\varepsilon^{^{\omega}}}{\varepsilon_{_{\Omega}}}\right)^{^2}.
\end{equation}
}
\end{lemma}
{\it Proof} \,\,\,Notice that in course  of  shrinking the  role
of the  resonance eigenvalue is  being  played by  various
eigenvalues of the  Schr\"{o}dinger operator on the  quantum well.
Generically the resonance eigenvalue is  simple and the operator
$K^{^{-1}}_{_{-}}\,\,{\cal K}_{_{0}}$ is a bounded operator at the
resonance  value of the spectral parameter. Since the DN-map is
homogeneous of degree $-1$ and $K^{^{\Lambda}}_{_-} = K_{_{-}}$
admits an estimate  by $\varepsilon_{_{\omega}}$ from above, see
(\ref{2}) below, the  estimate of the non-resonance term is
obvious if  the distance of the  corresponding  eigenvalue
$\lambda_{_{s}}$  from the Fermi level  $E_{\Lambda}$ remains
strictly positive in course of shrinking. This gives the required
estimate of the contribution from all non-resonance terms with
eigenvalues which do not approach the Fermi level $E_{\Lambda}$ in
the course of the shrinking. The number of others non-resonance
terms, which correspond to eigenvalues approaching  the  Fermi
level in course of shrinking, is finite. It is sufficient to
estimate the contribution from a  single  non-resonance term.
Consider the term which corresponds to the eigenvalue
$\frac{\lambda_1}{\varepsilon^{^2}_{_{\Omega}}}$
    closest to  the  resonance eigenvalue
$\lambda'_0 = \frac{\lambda_0}{\varepsilon^{^2}_{_{\Omega}}}
\approx \Lambda$. Notice first that the shrinking of the
normalized eigenfunction   $\varphi_{_{1}}$  of the  dimensionless
Schr\"{o}dinger  operator is described by the formula
\[
\varphi'_{_{1}} (x') = \frac{\varphi_{_{1}}
\left(\frac{x'}{\varepsilon_{_{\Omega}}}\right)}
{\left[\int_{_{\Omega'}}  |\varphi_{_{1}}|^{^2}
\left(\frac{x'}{\varepsilon_{_{\Omega}}}\right)
dx'\right]^{^{1/2}}} \approx \varphi_{_{1}}
\left(\frac{x'}{\varepsilon_{_{\Omega}}}\right) \,\,
\varepsilon^{^{-1}}_{_{\Omega}}.
\]
Then the  normal derivative of the  shrinking eigenfunction is
transformed as
\[
\frac{\partial \varphi'_{_{1}}}{\partial n} = \frac{\partial
\varphi_{_{1}}}{\partial n}\,\,
\frac{1}{\varepsilon^{^{2}}_{_{\Omega}}}.
\]
If the  Fermi-level $\Lambda$ is kept constant in  course of
shrinking, then the closest to the Fermi-level resonance
eigenvalue is shifted to the spectral  point  $\lambda'_0 \approx
 \Lambda \varepsilon^{^{2}}_{_{\Omega}}$  of the
 ``dimensionless''
operator, with $\lambda'_0 $ ~dimension [mass]$^{^{-1}}$. Then the
spacing $\rho'(\Lambda)$ on the resonance level is calculated as
$\frac{\rho \left( E_{_{\Lambda}}\varepsilon^{^{2}}_{_{\Omega}}
\right)}{\varepsilon^{^{2}}_{_{\Omega}}}$. The projection
$P_{_{E}}$ is presented via  multiplication by the indicator of
the bottom-section, and the  norm of it is homogeneous first
degree, hence proportional to  $\delta^{^{\omega}}$. Combining all
these facts we obtain the following estimate for the contribution
to  the  DN-map from on the resonance lever from a  single
non-resonance term
     \begin{equation}
     \label{1}
\parallel \frac{P_{_{E}}\frac{\partial \varphi'_{_{1}}}{\partial n}\rangle \,
\langle P_{_{E}}\frac{\partial \varphi'_{_{1}}}{\partial
n}}{\lambda'_{_1} - \lambda'_{_0}}
\parallel \leq \frac{\sup_{\Gamma}|\bigtriangledown \varphi |^2}
{\rho\left(\Lambda \varepsilon^{^{2}}_{_{\Omega}}\right)}\,\,
\frac{\varepsilon^{^{\omega}}}{\varepsilon^{^2}_{_{\Omega}}}.
     \end{equation}
This result is in full agreement with the fact that the  DN-map is
homogeneous of order $-1$. Then we  have  $\parallel {\cal
K}_{_{0}}
\parallel \leq C \frac{\varepsilon^{^{\omega}}}{\varepsilon^{^2}_{_{\Omega}}}$.
\par
One can easily obtain the  estimate for $K_{_{-}}(\lambda)$ at the
Fermi-level:
\begin{equation}
     \label{2}
\parallel \left(K_{_{-}}(\lambda)\right)^{^{-1}} \parallel \approx
\max_m \sqrt{\frac{\mu_m^{^{\bot}}}{ \mu_m^{^{\parallel}}}}\,\,
     \frac{\varepsilon_{_{\omega}}}{\pi }
\end{equation}
if  $2 \mu_m^{^{\bot}} \left(\varepsilon^{^{\omega}}\right)^{^2}
\max_{_m} |E_{_{\Lambda}} - V_m|  << \pi $. Summarizing the
estimates (\ref{1},\ref{2})  we obtain the announced statement.
\vskip0.5cm \noindent {\bf Remark} Introducing the  positive
operator $|K_{_{-}}|:= -K_{_{-}} $ one can derive  a  similar
estimate in symmetrized form:
\[
|| |K_{_{-}}|^{^{-1/2}} {\cal K}_{_0}|K_{_{-}}|^{^{-1/2}} || \leq
C_0
\left(\frac{\varepsilon^{^{\omega}}}{\varepsilon_{_{\Omega}}}\right)^{^2}
\]
with some  constant $C_0$.
\par
 The shrinking of the network, with the constant Fermi level
$\Lambda $, can be applied to each quantum well and each quantum
wire separately. Then the DN map on the joining of the wells is a
direct sum of the DN-maps of quantum wells $\Omega_t,\,\,
t=1,2,\dots$, which implies:
\[
    \parallel {\cal K}_{_{0}} \parallel <
\max_{_{s}}\frac{\sup_{\Gamma}|\bigtriangledown \varphi_s |^2}
{\rho\left(\Lambda \varepsilon^{^{2}}_{_{\Omega}}\right)}\,\,
\frac{\varepsilon^{^{\omega}}}{\varepsilon^{^2}_{_{\Omega}}},
\]
and
\begin{equation}
\label{thin}
\parallel |K_{_{-}}|^{^{-1/2}}{\cal K}_{_0}|K_{_{-}}|^{^{-1/2}}
 \parallel \leq \max_{_{s,m}}
\sqrt{\frac{\mu_m^{^{\bot}}}{ \mu_m^{^{\parallel}}}}
     \frac{\sup_{\Gamma}|\bigtriangledown \varphi_s |^2}
{\pi \rho\left(\Lambda \varepsilon^{^{2}}_{_{\Omega}}\right)}\,\,
\left(\frac{\varepsilon_{_{\omega}}} {\varepsilon_{_{\Omega}}}
\right)^{^{2}}:= C_0 \,\, \left(\frac{\varepsilon^{^{\omega}}}
{\varepsilon_{_{\Omega}}}\right)^{^{2}},
\end{equation}
with an  absolute constant  $C_0$ depending  on the
 shape of  the  network. If the  shrinking of
details of the  network (the finite and  infinite wires and vertex
domains) is independent, then max in the preceding formula is
taken over all contacting pairs of details $\Omega_t,\, \omega^m$
and the maximal value of the ratio appears  in the  right  side.
\par
{\bf Definition}  We say that  {\it the network is thin on  the
Fermi-level  $\Lambda$}, if the  condition  $\parallel
|K_{_{-}}|^{^{-1/2}}{\cal K}_{_0}|K_{_{-}}|^{^{-1/2}}\parallel < 1
$ is fulfilled  for resonance values of the spectral parameter
near to the  Fermi level: $\lambda \approx \Lambda $.
\par
This condition is obviously fulfilled for shrinking networks, if
$\frac{\varepsilon_{\omega}}{\varepsilon_{\Omega}} << 1 $. But
practically  for given network and  fixed  $\Lambda$ this
condition can be also verified sometimes, based on  direct
calculations. In particular for the quantum switch \cite{boston}
based a circular quantum well radius $1$ with $4$ quantum wires
width $\delta$ attached  to it (centered at the points $0,\,
\pm\frac{2\pi}{3},\, \pi $) the condition is fulfilled  if $\delta
< 1/2$, due  to presence of some ``natural"  small  parameter.
\par
 The following statement (Theorem 3.3) reveals the
structure of singularities in the above representation (\ref{DNr})
for thin (but  non necessarily shrinking) networks.
\par
To  formulate and  proof  the  statement we  need  more elaborated
notations.  Separate the resonance term in matrix elements of the
DN-map of the operator $L_0$ presented  as  a  matrix with respect
to the  basis  $E_{\pm}$:
\[
P^{^{\Lambda}}_+\left({\cal D}{\cal N}\right) P^{^{\Lambda}}_+ =
\frac{1}{\left(2 \mu_{_{\Omega}}\right)^{^{2}}}
\frac{{\phi}^{^{+}}_0\rangle \langle {\phi}^{^{+}}_0} {\lambda -
\lambda^0_0} + \sum_{s\neq 0}\frac{1}{\left(2
\mu_{_{\Omega}}\right)^{^{2}}} \frac{\hat{\phi}^{^{+}}_s\rangle
\langle \hat{\phi}^{^{+}}_s} {\lambda - \lambda^0_s} :=
\frac{1}{\left(2 \mu_{_{\Omega}}\right)^{^{2}}}
    \frac{{\phi}^{^{+}}_0\rangle \langle {\phi}^{^{+}}_0}
{\lambda - \lambda^0_0} + {{\cal K}}_{++}
\]
\[
P^{^{\Lambda}}_+\left({\cal D}{\cal N}\right) P^{^{\Lambda}}_- =
\frac{1}{\left(2 \mu_{_{\Omega}}\right)^{^{2}}}
\frac{{\phi}^{^{+}}_0\rangle \langle {\phi}^{^{-}}_0} {{\lambda} -
{\lambda_0}} + \sum_{s\neq 0} \frac{1}{\left(2
\mu_{_{\Omega}}\right)^{^{2}}} \frac{{\phi}^{^{+}}_s\rangle
\langle {\phi}^{^{-}}_s} {{\lambda} - {\lambda^0_s}} :=
\frac{1}{\left(2 \mu_{_{\Omega}}\right)^{^{2}}}
    \frac{{\phi}^{^{+}}_0\rangle \langle {\phi}^{^{-}}_0}
{{\lambda} - {\lambda^0_0}} + {{\cal K}}_{+-}
\]
\[
P^{^{\Lambda}}_- \left({\cal D}{\cal N}\right) P^{^{\Lambda}}_+ =
\frac{1}{\left(2 \mu_{_{\Omega}}\right)^{^{2}}}
\frac{{\phi}^{^{-}}_0\rangle \langle {\phi}^{^{+}}_0} {{\lambda} -
{\lambda^0_0}} + \sum_{s\neq 0}\frac{1}{\left(2
\mu_{_{\Omega}}\right)^{^{2}}} \frac{{\phi}^{^{-}}_s\rangle
\langle {\phi}^{^{+}}_s} {{\lambda} - {\lambda^0_s}}
:=\frac{1}{\left(2 \mu_{_{\Omega}}\right)^{^{2}}}
    \frac{{\phi}^{^{-}}_0\rangle \langle {\phi}^{^{+}}_0}
{{\lambda} - {\lambda^0_0}} + {{\cal K}}_{-+}
\]
with  ${{\cal K}}_{- +} = \left( {{\cal K}}_{+ -} \right)^{^+} $,
and
\[
P^{^{\Lambda}}_-\left({\cal D}{\cal N}\right) P^{^{\Lambda}}_- =
\frac{1}{\left(2 \mu_{_{\Omega}}\right)^{^{2}}}
\frac{{\phi}^{^{-}}_0\rangle \langle {\phi}^{^{-}}_0} {{\lambda} -
{\lambda^0_0}} + \sum_{s\neq 0} \frac{{\phi}^{^{-}}_s\rangle
\langle {\phi}^{^{-}}_s} {{\lambda} - {\lambda^0_s}}
:=\frac{1}{\left(2 \mu_{_{\Omega}}\right)^{^{2}}}
    \frac{{\phi}^{^{-}}_0\rangle \langle {\phi}^{^{-}}_0}
{{\lambda} - {\lambda^0_0}} + {{\cal K}}_{--}.
\]
Here  ${\cal K}_{_{++}},\,{\cal K}_{_{\pm}},\,{\cal K}_{_{--}}\, $
are  the matrix  elements of the contribution ${\cal K}_{_{0}}$ to
the DN-map from the non-resonance eigenvalues. The  operators
${\cal K}_{_{++}},\,{\cal K}_{_{+-}}$  are  bounded  in
$W_2^{3/2-0} (\Gamma)$,\, ${\cal K}_{_{-+}}= {\cal K}_{_{+-}}^+$
for  real $\lambda$ and  ${\cal K}_{_{--}}$ acts from $W_2^{3/2-0}
(\Gamma)$ into $W_2^{1/2-0} (\Gamma)$ the  same  way as $K_-$
does. Then the expression  (\ref{DNr}) may be written  as
\[
{\cal D}{\cal N}^{^{\Lambda}} =\frac{1}{\left(2
\mu_{_{\Omega}}\right)^{^{2}}} \frac{{\phi}^{^{+}}_0\rangle
\langle {\phi}^{^{+}}_0} {{\lambda} - {\lambda^0_0}} + {{\cal
K}}_{+-} \,\,+
\]
\begin{equation}
\label{DNapprox} \left[\frac{1}{\left(2
\mu_{_{\Omega}}\right)^{^{2}}} \frac{{\phi}^{^{+}}_0\rangle
\langle {\phi}^{^{-}}_0} {{\lambda} - {\lambda^0_0}} + {{\cal
K}}_{+-}\right] \frac{I}{ \frac{1}{\left(2
\mu_{_{\Omega}}\right)^{^{2}}} \frac{{\phi}^{^{-}}_0\rangle
\langle {\phi}^{^{-}}_0} {{\lambda} - {\lambda^0_0}} + {{\cal
K}}_{--} - K_-}
    \left[\frac{1}{\left(2 \mu_{_{\Omega}}\right)^{^{2}}}
    \frac{{\phi}^{^{-}}_0\rangle \langle {\phi}^{^{+}}_0}
{{\lambda} - {\lambda^0_0}} + {{\cal K}}_{-+}\right].
\end{equation}
The positive operator $-K_{_{-}} = |K_{_{-}}|$  can be  estimated
from below by the  distance from $\lambda$ to the  lowest upper
threshold
    \[
\rho_{_{-}}(\lambda) = \mbox{min}_{_{_{\frac{\pi^2 l^2} {2
\mu^{\bot}_{_m}\delta_m^2} + V_{_m} > \Lambda}}}
    \,\,\,\,\sqrt{2\mu^{^{\parallel}}_m}\,\,\,
    \sqrt { \frac{\pi^2 l^2}{2 \mu^{\bot}_{_m}\delta_m^2} + V_{_m} - \lambda},
\]
\[
\displaystyle \langle -K_{_{-}} u,\,u \rangle \geq \rho_{_{-}}
\parallel u
\parallel^{^2}_{_{L_2 (\Gamma)}}.
\]
\begin{theorem}{\it If the  network  is  thin  on the Fermi-level
$\Lambda$, then the pole  of  the  DN-map  ${\cal D}{\cal N}$ at
the simple resonance eigenvalue $\lambda^0_0$ of the  operator
$L_0$ on the compact part $\Omega_0$ of the network (the
singularity of the first addendum of (\ref{DNr})) is compensated
by the pole of the second addendum and disappears as a singularity
of  the  whole function ${\cal D}{\cal N}^{^{\Lambda}}$ so  that
the whole expression (\ref{DNr}) is regular at the point
$\lambda^0_0$. A new pole appears as a closest to $\Lambda$ zero
eigenvalue  of the denominator $\left[P^{^{\Lambda}}_- \left({\cal
D}{\cal N}\right) P^{^{\Lambda}}_- - K_{_{-}}\right] \nu = 0 $. }
\end{theorem}
{\it Proof} \,\,\, If the  network is  thin, $\parallel
|K_{_{-}}|^{^{-1/2}}{\cal K}_{_0} (\lambda_0)|K_{_{-}}|^{^{-1/2}}
\parallel < 1 $, then the operator
\[
{\cal K}_{_{--}} - K_{_{-}} := k
\]
is  invertible:  $k^{^{_1}} =|K_{_{-}}|^{^{-1/2}}
    \left(1 - |K_{_{-}}|^{^{-1/2}} {\cal K}_{_{--}}|K_{_{-}}|^{^{-1/2}}\right)
     |K_{_{-}}|^{^{-1/2}}$.
Then the  middle  term  of  the  above product (\ref{DNapprox})
can be found  as a  solution of  the  equation
\[
\left[ \frac{1}{\left(2 \mu_{_{\Omega}}\right)^{^{2}}}
\frac{{\phi}^{^{-}}_0\rangle \langle {\phi}^{^{-}}_0} {{\lambda} -
{\lambda^0_0}} + {{\cal K}}_{--} - K_{_{-}}  \right] u = f,
\]
\[
u = k^{^{-1}}\,\,f - \frac{1}{\cal D}\,\,\, k^{^{-1}}\, \,
{\phi}^{^{-}}_{0}\rangle\,\,\langle {\phi}^{^{-}}_{0},\,
k^{^{-1}}\,\, f \rangle,
\]
where ${\cal D} =\left(2
\mu_{_{\Omega}}\right)^{^{2}}\,\,\left({\lambda} - {\lambda}^0_0
\right) + \langle {\phi}^{^{-}}_{0},\,
    k^{^{-1}}\,\, \, {\phi}^{^{-}}_{0}\rangle$. Zeroes of the
function $D$ coincide with singularities of the middle  term in
the above  formula (\ref{DNr}) for the  ${\cal D}{\cal
N}^{^{\Lambda}}$ and  hence coincide  with the  eigenvalues of the
intermediate operator $L_{_{\Lambda}}$.
\par
Substituting that expression into (\ref{DNapprox}) we notice that
all polar terms containing  the factors $\left({\lambda}-
{\lambda^0_0}\right)^{^{-1}}$ compensate each  other so  that the
sum of them vanishes :
\[
{\cal D}{\cal N}^{^{\Lambda}} = \frac{1}{\left(2
\mu_{_{\Omega}}\right)^{^{2}}} \frac{{\phi}^{^{+}}\rangle\,\langle
{\phi}^{^{+}}}{{\lambda}- {\lambda}_0} \left[1 - \frac{\langle
{\phi}^{^{-}},\,k^{^{-1}}\,\, {\phi}^{^{-}} \rangle} {{\lambda}-
{\lambda}_0} + \frac{1}{\left(2 \mu_{_{\Omega}}\right)^{^{2}}}
\frac{\langle {\phi}^{^{-}},\,k^{^{-1}}\,\,{\phi}^{^{-}}
\rangle^{^{2}}} {{\cal} D\left(\hat{\lambda}- {\lambda}_0\right)}
\right] -
\]
\[ \frac{1}{\left(2 \mu_{_{\Omega}}\right)^{^{2}}}
\frac{{\cal K}_{+ -} k^{^{-1}}\,{\phi}^{^{-}}\rangle\, \langle
{\phi}^{^{+}} } {{\lambda}- {\lambda}_0}\left[ I - \frac{\langle
{\phi}^{^{-}},\,k^{^{-1}}\,\, {\phi}^{^{-}} \rangle} {{\cal}
D}\right]-
\]
\[
\frac{1}{\left(2 \mu_{_{\Omega}}\right)^{^{2}}}
\frac{{\phi}^{^{+}} \rangle\, \langle {\cal K}_{+ -} k^{^{-1}}\,\,
{\phi}^{^{-}}} {{\lambda}- {\lambda}_0} \left[ I - \frac{\langle
{\phi}^{^{-}},\,k^{^{-1}}\,\, {\phi}^{^{-}} \rangle} {{\cal}
D}\right] +
\]
\[
\frac{1}{\left(2 \mu_{_{\Omega}}\right)^{^{2}}} \frac{{\cal K}_{+
-}k^{^{-1}}\,\, {\phi}^{^{-}}\rangle\,
    \langle {\cal K}_{+ -} k^{^{-1}}\,\,
{\phi}^{^{-}} }{{\cal D}} + \dots = \]
\begin{equation}
\label{second}
 \frac{1}{\left(2 \mu_{_{\Omega}}\right)^{^{2}}}
\frac{{\phi}^{^+} -{\cal K}_{+ -}
k^{^{-1}}\,\,{\phi}^{^{-}}\rangle\,\,
    \langle {\phi}^{^+} - {\cal K}_{+ -} k^{^{-1}}\,\,
{\phi}^{^{-}}}{\cal D} + \dots,
\end{equation}
where the  dots  stay for   terms defining the  regular summand of
${\cal D}{\cal N}^{^{\Lambda}}$ in a  small neighborhood of the
resonance.
 \vskip0.5cm
\noindent {\bf Remark 1} One can see that on the first step of the
approximation procedure we obtain the pole of ${\cal D}{\cal
N}^{^{\Lambda}}$ at the simple zero of the denominator ${\cal D}$
with the same residue ${\phi}^{^+}\rangle \,\, \langle
{\phi}^{^+}$ as  ${\cal D}{\cal N}$, in full agreement with
physical folklore. The residue, with a small correction
    obtained on the  second  step of  the  approximation procedure, is
given by
    \[
{\phi}^{^+} -{\cal K}_{+ -}k\,\,{\phi}^{^{-}}\rangle\,\,\langle
{\phi}^{^+} - {\cal K}_{+ -} k\,\, {\phi}^{^{-}}.
    \]
In particular this means that the portions $P_+ \frac{\partial
\varphi^r_0}{\partial n}$ of resonance eigenfunctions of  the
intermediate operator $L{_{\Lambda}}$ in the entrance subspace can
be  found via the successive approximation procedure (on the
second  step) as:
\[
    P_+ \frac{\partial \varphi^r_0}{\partial n} \approx
    {\phi}^{^+} - {\cal K}_{+ -}\,\, k\,\,{\phi}^{^{-}},
\]
if the  network is  thin. Similarly the shift of the  resonance
eigenvalue $\lambda^{^0}_{_0}$ to  the zero
$\lambda^{^{D}}_{_{0}}$ of the  denominator $D$
\[
D(\lambda^{^{D}}_{_{0}}) = 0
\]
can be  estimated in first order of the  approximation procedure
as
\[
\lambda^{^{D}}_{_{0}} = \lambda^{^{0}}_{_{0}} -
\frac{1}{(2\mu_{_{\Omega}})^{^2}}\,\, \langle \varphi_{_0},
|K_{_{-}}|^{^{-1}}\varphi_{_0}\rangle + \dots,
\]
where the  dots  stay for terms  estimated  by powers  of
\[
\frac{1}{(2\mu_{_{\Omega}})^{^2}}\,\, \langle \varphi_{_0},
|K_{_{-}}|^{^{- 1/2}}|{\cal K}_{_{--}}||K_{_{-}}|^{^{- 1/2
}}\varphi_{_0}\rangle
\]
 \vskip0.5cm
 {\bf Remark 2}
Note that  the  exponent  $K_{_{-}}$  can be estimated  from below
in terms of the wave-number  of the electron in the wires at the
resonance energy $\lambda_{_0}^{^0}$
\[
p_{_{_{min}}} (\lambda^{^0}_{_0}) = \mbox{min}_{_{_{\frac{\pi^2
l^2}{2 \mu^{\bot}_{_m}\delta_m^2} + V_{_m} < \Lambda}}}\,\,\,
\sqrt{2 \mu^{^{\parallel}}_{_{m}}}
    \sqrt {\lambda_0 - \frac{\pi^2 l^2}{2 \mu^{\bot}_{_m}\delta_m^2} -
    V_{_m}}.
\]
In case of  thin networks  the wave-number exceeds the
contribution to the matrix element ${\cal D}{\cal N}_{_{--}}$ of
the  DN-map of $L_0$ from the  neighboring non-resonance
eigenvalues.

The eigenvalues and the eigenfunctions of the  discrete spectrum
of  the intermediate operator  $L_{_{\Lambda}}$ can be  found
either by the minimizing of  the  corresponding Rayleigh ratio, or
from the corresponding dispersion equation,
\[
P^{^{\Lambda}}_- \left({\cal D}{\cal N}\right) P^{^{\Lambda}}_- -
K_{-} \nu = 0
\]
 involving the DN-map of  the Schr\"{o}dinger operator $L_{0}$
with  zero boundary condition on the  border of the compact part
$\Omega_0$ of the  network.
     \begin{theorem}{\it The eigenvalues of the operator $L_{_{\Lambda}}$
may be found as vector zeroes $(\lambda,\, \nu_{_{\lambda}})$ of
the dispersion equation
\[
{\bf D}(\lambda)\nu_{_{\lambda}} = {\cal D}{\cal N}_{--}
(\lambda)\,\, \nu_{_{\lambda}} - K_{_{-}}(\lambda)\,\,
\nu_{_{\lambda}} = 0.
\]
In particular for  values  of  the  spectral parameter  between
the maximal  lower threshold $\lambda^{^{\Lambda}}_{_{max}}$ and
minimal upper threshold $\lambda^{^{\Lambda}}_{_{min}}$ the
dispersion equation  takes the form
\begin{equation}
\label{dispers} \nu = {|K_{-}|}^{^{-{1/2}}}\,\,{\cal D}{\cal
N}_{--}\,\,{|K_{-}|}^{^{-{1/2}}} \nu,
\end{equation}
with the  bounded  operator-functions $|K_{-}|^{^{-1/2}}\,\,{\cal
D}{\cal N}_{--}\,\,|K_{-}|^{^{-1/2}}$ in $L_2 (\Gamma)$. It may be
transformed  to an equation
\[
\nu - {\cal D} (H,\,\lambda) \nu  = 0,
\]
with a  trace class operator
\[{\cal
D} (H,\,\lambda) := (H + \lambda)^{^{2}}\,\,{\bf k}^{^{-1/2}}\,
{\cal P}^{^{+}}_{_{H}}{\cal R}_{_{\lambda}}{\cal P}_{_{H}}{\bf
k}^{^{-1/2}},
\]
where
\[
{\bf k} =  \left[|K_{_{-}}| + {\cal D}{\cal N}_{--}\,\,(-H) - (H +
\lambda){\cal P}^{^{+}}_{_{H}} {\cal P}_{_{H}}\right] =
\]
\[
(|K_{_{-}}|)^{^{1/2}} \left[I + (|K_{_{-}}|)^{^{-1/2}}\left({\cal
D}{\cal N}_{--}\,\,(-H) - (H + \lambda){\cal P}^{^{+}}_{_{H}}
{\cal P}_{_{H}}
\right)(|K_{_{-}}|)^{^{-1/2}}\right](|K_{_{-}}|)^{^{1/2}}
\]
and   ${\cal R}_{_{\lambda}},\,\,{\cal P}_{_{\lambda}}$ are
respectively the resolvent and the Poisson map of the operator $L$
on the compact part of the network with zero boundary condition.
The corresponding scalar  equation may be  presented in the form
\begin{equation}
\label{Dispersq} \det \left[ I - {\cal D} (H,\,\lambda)\right] =
0.
\end{equation}
}
\end{theorem}
{\it Proof}\,\,\,\,The projections $u_{_{_{\Gamma}}} \in E_{_{-}}$
of the eigen-function $u,\,\, L_{_{\Lambda}} u = \lambda u $ of
the operator $L_{_{\Lambda}}$ onto the cross-sections $\Gamma$ of
the open  channels  should fulfill  the  condition
    $\frac{\partial u}{\partial n} \bigg|_{_{\Gamma}} =
K_{_{-}}^{^{\Lambda}} (\lambda) u_{_{\Gamma}}$.
    On the  other hand  the  restrictions of $u$ onto  the  compact
part of the  network fulfills  the  corresponding  homogeneous
equation, hence
    $\frac{\partial u}{\partial n} \bigg|_{_{\Gamma}} = {\cal D}{\cal
N}_{--}\,\,(\lambda) $. Matching  both data with  the boundary
conditions (\ref{bcond}) we obtain the dispersion equation:
    \[
 \,\,\, K_{_{-}}^{^{\Lambda}}
(\lambda) u_{_{\Gamma}} = \,\,\, P^{^{\Lambda}}_{_{-}}{\cal
D}{\cal N}\,\,P^{^{\Lambda}}_{_{-}}\,\,(\lambda)\,\,
u_{_{\Gamma}}.
\]
Due to the invertibility  of  $ K_{_{-}}^{^{\Lambda}} (\lambda)$
below $\lambda^{^{\Lambda}}_{_{min}}$ we  obtain the  first
statement (\ref{dispers}) of  the  theorem. The  second statement
requires the iterated  Hilbert identity  for  the  DN-map, see
\cite{DN01}:
\[
{\cal D}{\cal N}\,\,(\lambda) = {\cal D}{\cal N}\,\,(-H) - (H +
\lambda){\cal P}^{^{+}}_{_{H}} {\cal P}_{_{H}} - (H +
\lambda)^{^{2}}{\cal P}^{^{+}}_{_{H}} R_{_{\lambda}} {\cal
P}_{_{H}}.
\]
 Both second and third terms
of the sum in the right hand side are compact operators in $L_2
(\Gamma)$, see the remark after the  theorem 3.2, if the  compact
part $\Omega_0$ of the network has a piece-wise smooth boundary
with Meixner boundary conditions at the inner corners. Moreover,
the third term, after framing by  factors  $K_{_{-}}^{^{\Lambda}}$
 is an operator
with a finite  trace. Then the operator ${\cal D} (H,\,\lambda)$
has a finite trace  too and  the dispersion equation may be
presented in  form
\[
\det \left[ I - {\cal D} (H,\,\lambda)\right] = 0.
\]
\vskip0.5cm

\section{Scattering matrix}
 In this  section we  rephrase  some   results  of  previous  section in
 terms  of  Scattering  Matrix and then  derive  and  interpret the
corresponding  ``one-pole  approximation''.
 \par
 Components of the  Scattered  waves of the
Schr\"{o}dinger operator ${\cal L}$  in the lower channels on
semi-infinite wires $\omega^m $ are presented  by linear
combination of modes: oscillating exponentials combined with
eigenfunctions of cross-sections in open channels and decreasing
exponentials combined  with corresponding eigenfunctions of
cross-sections in closed channels:
\[
\psi_{+} = \sum_{\frac{\pi^2 l^2}{2 \mu_m^{\bot}\delta_m} + V_m<
\Lambda} e_m^l \,\, \nu^m_l e^{-i\sqrt{2
\mu_m^{\parallel}}\sqrt{\lambda - V_m \frac{\pi^2 l^2}{2
\mu_m^{\bot}\delta_m}}\,\,x}+
\]
\[
\sum_{\frac{\pi^2 k^2}{2 \mu_m^{\bot}\delta_m} + V_m <\Lambda}
e^m_l\,\,S^{l,k}_{m,n}\nu_k^n\,\,
    e^{i\sqrt{2 \mu_m^{\parallel}}\sqrt{\lambda - V_m -\frac{ \pi^2
l^2}{2 \mu_m^{\bot}\delta_m}}\,\,x}+
\]
\begin{equation}
\label{SAnsatz} \sum_{\frac{\pi^2 k^2}{2 \mu_m^{\bot}\delta_m} +
V_m > \Lambda} e^m_l\,\,s^{l,k}_{m,n}\nu_k^n\,\,
    e^{-\sqrt{2 \mu_m^{\parallel}}\sqrt{ V_m + \frac{ \pi^2 l^2}{2
\mu_m^{\bot}\delta_m} - \lambda}\,\,x},
\end{equation}
where $\nu^{^m}_{_{l}}$ are the components of the  incoming plane
wave, and the finite  matrix  $S = \left\{ S^{l,k}_{m,n}\right\}$
is a  Scattering  matrix---the main object  of  our  search. We
present the  above  Scattering Ansatz in the following short form
(\ref{Sanzatz}) introducing the notations:
    $\sum_{ml} e^m_l \,\, \nu_m^l = \nu_+ \in E_+,\,
\, \left\{ S^{l,k}_{m,n}\right\} = {\bf S},\,\, \left\{
s^{l,k}_{m,n}\right\} = {\bf s}$ and  the  diagonal matrices in
$E^{\Lambda}_+,\,E^{\Lambda}_- $:
\[
K_{_{+}}= \left\{ \sqrt{2 \mu_m^{\parallel}}\,\,\sqrt{\lambda -
V_m -  \frac{\pi^2 l^2}{2 \mu_m^{\bot}\delta_m}}\right\}
\]
for $\frac{\pi^2 k^2}{2 \mu_m^{\bot}\delta_m} + V_m <\Lambda$ and
\[
K_{_{-}}= - \left\{\sqrt{2 \mu_m^{\parallel}}\sqrt{ V_m + \frac{
\pi^2 l^2}{2 \mu_m^{\bot}\delta_m} - \lambda}\right\}
\]
for  $\frac{\pi^2 k^2}{2 \mu_m^{\bot}\delta_m} + V_m > \Lambda$,
then the  Scattering Ansatz in the  wires  is:
\begin{equation}
\label{Sanzatz} \psi_{+} :=  e^{^{-i  K_{_{+}}x}}\nu_{_{+}} +
e^{^{ i  K_{_{+}}x}} {\bf S}\nu_{_{+}} + e^{^{ K_{_{-}}x}}{\bf
s}\nu_{_{+}}.
\end{equation}
To derive the formula for the Scattering matrix of the operator
${\cal L}$ on the whole network $\Omega$ we  should substitute the
above Ansatz into the boundary conditions of matching  with
components  $\psi_{_{\Omega}}$ of the  Scattering Ansatz on the
vertex domains
\begin{equation}
\label{matching} P_{\pm}\left[ \psi \right]|_{\Gamma}= 0,\,
\,P_{\pm}\left[\frac{\partial \psi }{\partial n}\right]|_{\Gamma}=
0,
\end{equation}
where  $\left[ \psi \right]|_{\Gamma} = \left[\psi^{\omega}-
\psi_{_{\Omega}}\right]|_{\Gamma}$ and $\left[\frac{\partial \psi
}{\partial n}\right]|_{\Gamma} =
\left[\frac{1}{\mu^{^{\parallel}}}\frac{\partial
\psi^{^{\omega}}}{\partial n} - \frac{1}{{\bf \mu}_{_{\Omega}}}
\frac{\partial \psi_{_{\Omega}}}{\partial
n}\right]\bigg|_{\Gamma}$. The  components
$\left\{\psi^{\omega},\, \frac{\partial \psi^{^{\omega}}}{\partial
n} \right\}$ of the above Ansatz in the wires are obtained by
straightforward differentiation. However, the corresponding
connection between the values of the components
$\left\{\psi_{_{\Omega}},\,\frac{\partial
\psi_{_{\Omega}}}{\partial n} \right\}$ of the Ansatz in the
vertex domains $\Omega_t$ is  given  by the DN-map:
\[
{\cal D}{\cal N} \psi_{_{\Omega}}\bigg|_{_\Gamma}
=\frac{1}{2\mu_{_{\Omega}}}\,
    \frac{\partial \psi_{_{\Omega}}}{\partial n}\, \bigg|_{_{\Gamma}}.
\]
Combining these  connections and  using the boundary  conditions
(\ref{matching}), we obtain an equation for the  undefined
coefficients of  the Ansatz (\ref{SAnsatz}).
\par
In this section we follow this standard program assuming that  the
network consists of ``extended'' vertex domains constituting the
compact part $\Omega_0$, see section 1, and several  semi-infinite
wires $\omega^{^m},\, m=1,2,\dots$ width $\delta_m$ attached
orthogonally to $\Omega_0$ at  the bottom sections $\gamma_m$,
$\Gamma = \cup_{m} \gamma_m$. The potential $V (x)$ of the
Schr\"{o}dinger
    operator
\[
L u = -\bigtriangleup u + V(x) u
\]
takes constant values $V_m$ on the  wires $\omega^{^m}$ and  is a
real bounded measurable function on  vertex  domains. Assume that
the  DN-map of  the operator $l_0$ on each vertex domain , and
hence on  $\Omega_{_{0}}$
 is already constructed and presented  (formally)
    by the  spectral  series in terms of the eigenfunctions  $\varphi_r$ and
eigenvalues $\lambda_r$
    corresponding  to  zero  boundary conditions on  $\partial \Omega_0$:
\begin{equation}
\label{DN} {\cal D}{\cal N}(x,x') =  \sum_s \frac{\frac{ \partial
\varphi_r (x)}{\partial n_x}\rangle\, \langle  \frac{\partial
\varphi_r (s)}{\partial n_{x'}}} {\lambda - \lambda^0_r} .
\end{equation}
Here  ${\cal D}{\cal N}$  is  a direct sum of  DN-maps of  vertex
domains. As a  result of  this  steps, we  obtain below  the
explicit formula for  Scattering matrix  in terms  of  matrix
 elements  of  ${\cal D}{\cal N}$  with respect  to  the
 orthogonal decomposition  of  the  entrance  subspace
 $E = E_{_{+}} \oplus E_{_{-}}$, see  (\ref{Scatt}).
\par
On the  other  hand, choosing  the  Fermi  level $\Lambda$ we  can
define the intermediate operator $L_{\Lambda}$ and  construct the
corresponding  DN-map  ${\cal D}{\cal N}^{^\Lambda}$ as described
in the previous  section. This  object  is  more  sophisticate,
because  it  already contains  an  information  on closed
channels, where  the  partial  matching  is  already achieved.
Really  in   second  case  we calculate Scattering solutions of
the operator $L$ matching limit  values on real axis of the
spectral parameter $\lambda$  of square-integrable solutions of
homogeneous intermediate equations with  boundary data from
$E_{_{+}}$ on $\Gamma$ - with  Scattering Ansatz  in open challels
only:
\begin{equation}
\label{Sansatz_open} \psi^{\omega} = e^{-i K_+ x} \nu  + e^{
iK_{+}x} {\bf S} \nu.
\end{equation}
The boundary  data of the  solution  of the  intermediate equation
are connected via the corresponding  (intermediate) DN-map:
\[
{\cal D}{\cal N}^{^{\Lambda}}\psi_0 \bigg|_{\Gamma} = \frac{1}{ 2
\mu_m^{\parallel}} P_{_{+}} \frac{\partial \psi_0}{\partial
n}\bigg|_{\Gamma}.
\]
Assuming that $\nu = \sum_{m,l}\nu_{ml}e^m_l \in E_{_+} $ we
obtain from the  matching conditions the following
  equation:
\[
 {\cal D}{\cal N}^{^{\Lambda}}
\left[ I + S \right] = \left[ - i K_{+} +  i K_{+} S \right]
\]
and  the  corresponding representation  for  the  Scattering
matrix, see  below (\ref{Smatrix}). Summarizing  two approaches to
calculation of the  Scattering  matrix, described  above, we
obtain the  following  statement:
\begin{theorem}{
The  following  two  formulae  are  valid  for the  Scattering
matrix on the  Network:
\par
\noindent a. The formula in terms of the standard DN-map ${\cal
D}{\cal N}$   of the operator $L_0$ on the compact part $\Omega_0$
of the  network:
\begin{equation}
\label{Scatt} S(\lambda)= - \frac{ \left({\cal D}{\cal
N}\right)_{++} -
    \left({\cal D}{\cal N}\right)_{+-}
\frac{I}{\left({\cal D}{\cal N}\right)_{--} - K_{-}}\left( {\cal
D}{\cal N} \right)_{-+} + iK_{_{+}}} {\left({\cal D}{\cal
N}\right)_{++} -
    \left({\cal D}{\cal N}\right)_{+-}
\frac{I}{\left({\cal D}{\cal N}\right)_{--} - K_{-}}\left( {\cal
D}{\cal N} \right)_{-+} - iK_{_{+}}},
\end{equation}
and
\par
\noindent b. The formula for  Scattering  matrix
 in terms  of  the  DN-map  ${\cal D}{\cal N}^{^{\Lambda}}$ of
the  intermediate operator:
\begin{equation}
\label{Smatrix} {\bf S} (\lambda) = - \frac{{{\cal D}{\cal
N}}^{^{\Lambda}} + i K_+}{{{\cal D}{\cal N}}^{^{\Lambda}} - i
K_+}.
\end{equation}
In both formulae  the  denominator is the  first  factor of the
product. }
\end{theorem}
\par
Note  that both above  formulae  are  equivalent  due  to
connection  between  the  DN-maps of  the  intermediate operator
and  DN-map  of  Dirichlet problem  on the  compact part  of the
network, established in previous  section.  Though the formula
(\ref{Smatrix}) contains more sophisticate object ${\cal D}{\cal
N}^{^{\Lambda}}$ compared with ${\cal D}{\cal N}$, it may be  more
convenient for calculating resonances, see \cite{Lax}, and for the
description of transition processes, because  the Scattering
matrix  is presented as a combination of  bounded  operators.
\par
One  can  notice  that leading terms  in the  numerator  and
denominator  of  the above  expression   (\ref{Smatrix}) for  the
Scattering matrix near the  resonance eigenvalue $\lambda_{_{0}}$
are : the polar term
\[
\frac{P_{+}\frac{ \partial \varphi_{_0} (x)}{\partial n_x}\rangle
\,\,\langle P_{+}\frac{
\partial \varphi_{_0} (s)}{\partial n_s}} {\lambda -
\lambda^{^{\Lambda}}_{_{0}} } : = \frac{ \phi_{_0}
(x)\rangle\,\,\langle \phi_{_0} (s)} {\lambda -
\lambda^{^{\Lambda}}_{_{0}} }
\]
and  $i K_{_{+}} (\lambda)$, which  is  estimated   from below on
the  spectral band  $\Delta_{_{\Lambda}} =
\left[\lambda^{^{\Lambda}}_{_{max}},\,\lambda^{^{\Lambda}}_{_{min}}\right]$:
\begin{equation}
\label{estim_K} \bigg| K^{^{-1}}_{_{+}} (\lambda)\bigg|\leq
\frac{1}{ \sqrt{2 \mu_m^{\parallel}}\,\,\sqrt{\lambda -
\lambda^{^{\Lambda}}_{_{max}}}}: = \frac{1}{\bf k}.
\end{equation}
Assume that the remaining non-resonance  part of DN-map
\[
\left[ {\cal D}{\cal N}^{^{\Lambda}}  - \frac{P_{+}\frac{
\partial \varphi_{_0} (x)}{\partial n_x}\rangle \,\,\langle
P_{+}\frac{
\partial \varphi_{_0} (s)}{\partial n_s}} {\lambda -
\lambda^{^{\Lambda}}_{_{0}}} \right] := {\cal D}{\cal
N}^{^{\Lambda}}_{_{0}}
\]
is subordinated  to  $i K_{_{+}}$ in a  small real neighborhood
$(\lambda_{_0} - \varepsilon,\, \lambda_{_0} + \varepsilon ): =
\Delta_{_{\varepsilon}}$ of the resonance eigenvalue, that is :
the  operator $K_{_{+}}^{^{-1/2}}\,{\cal D}{\cal
N}^{^{\Lambda}}_{_{0}}\, K_{_{+}}^{^{-1/2}} = {\bf K}$ is
estimated  as
\begin{equation}
\label{subord}
 |{\bf K}| = \bigg|K_{_{+}}^{^{-1/2}}\,{\cal D}{\cal
N}^{^{\Lambda}}_{_{0}}\, K_{_{+}}^{^{-1/2}} \bigg| \leq  d << 1,
\end{equation}
on thin or  shrinking network, with $d = d (\varepsilon)$. This,
in particular, means that the spacing on the resonance level is
large comparing with $\varepsilon$. Note that, due to analyticity
of both functions $K_{_{+}},\,{\cal D}{\cal
N}^{^{\Lambda}}_{_{0}}$ at the resonance level the above
subordination condition (\ref{subord}) can be extended to some
complex  neighborhood $U^{^C}_{_{\varepsilon}}$ of the resonance
eigenvalue $\lambda^{^{\Lambda}}_{_{0}}$
\par
Form  the unitary on real axis $\lambda$ combination of  the
leading terms in the numerator and  denominator
\begin{equation}
\label{ONEPOLE} S_{_{_0}} (\lambda) = \frac{i K_{_{+}} + \frac{
\phi_{_0} \rangle\,\,\langle \phi_{_0} } {\lambda -
\lambda^{^{\Lambda}}_{_{0}} }}{i K_{_{+}} - \frac{ \phi_{_0}
\rangle\,\,\langle \phi_{_0} } {\lambda -
\lambda^{^{\Lambda}}_{_{0}} }}.
\end{equation}
Under certain conditions this  combination may serve a ``one-pole
approximation '' of the Scattering matrix near the  resonance
eigenvalue. Zeroes of $S_{_{_0}}$ sit in upper half-plane  and can
be found from the algebraic equation  $ \lambda =
\lambda^{^{\Lambda}}_{_{0}} + i \langle
K^{^{-1}}_{_{+}}(\lambda)\,\phi_{_{0}},\,\phi_{_{0}}\rangle $. If
\begin{equation}
\label{dominant} |\langle K_{_{+}} \phi_{_{0}}, \phi_{_{0}}\rangle
| <  \varepsilon_{_1},\,\, \mbox{and} \,\,\,|\langle K'_{_{+}}
\phi_{_{0}}, \phi_{_{0}}\rangle| \leq q,\,\,
\frac{\varepsilon_{_1}}{1- q} < 1,
\end{equation}
then a single zero of it sits in the above complex neighborhood
$U^{^C}_{_{\varepsilon}}$ and it  can be  found  by  successive
approximations procedure.
\begin{theorem}{If both  conditions (\ref{subord},\ref{dominant}) are  fulfilled,
with  $d$ small  enough, then on  the some  real neighborhood
$U^{^{R}}_{_{\varepsilon}}$ of the resonance  eigenvalue  the
following estimate is  true
\begin{equation}
\label{one_appr} \sup_{_{\lambda \in }}\bigg| S(\lambda) -
S_{_{0}}(\lambda)\bigg|_{_{U^{^{R}}_{_{\varepsilon}}}} \leq
 \frac{2 d}{1-d} \,\frac{3}{2}\,
|K^{^{1/2}}_{_{+}}(\lambda^{^{\Lambda}}_{_{0}})|\,\,\,
 |K^{^{- 1/2}}_{_{+}}(\lambda^{^{\Lambda}}_{_{0}})|,
\end{equation}
and  there exist a  singe  zero of a  Scattering matrix  near to
the  above  zero  of  a  one-pole  approximation. }
\end{theorem}
{\it Proof}. With  notations  introduced  above  the  expression
(\ref{Smatrix}) can be  presented on $\Delta_{_{\varepsilon}}$ as
\[
 S_{_{_0}} (\lambda) =
\frac{i K_{_{+}} + \frac{ \phi_{_0} \rangle\,\,\langle \phi_{_0}}
{\lambda - \lambda^{^{\Lambda}}_{_{0}} } + {\cal D}{\cal
N}^{^{\Lambda}}_{_{0}} }{i K_{_{+}} - \frac{ \phi_{_0}
\rangle\,\,\langle \phi_{_0}}{\lambda -
\lambda^{^{\Lambda}}_{_{0}}} - {\cal D}{\cal
N}^{^{\Lambda}}_{_{0}}} =
\]
\begin{equation}
\label{Sapprox2} = K^{^{-1/2}}_{_{+}} \frac{iI_{_{+}} +
\frac{K^{^{-1/2}}_{_{+}} \phi_{_0} \rangle\,\,\langle
K^{^{-1/2}}_{_{+}}\phi_{_0}}{\lambda - \lambda^{^{\Lambda}}_{_{0}}
} + {\bf K}}{iI_{_{+}} - \frac{K^{^{-1/2}}_{_{+}} \phi_{_0}
\rangle\,\,\langle K^{^{-1/2}}_{_{+}}\phi_{_0}}{\lambda -
\lambda^{^{\Lambda}}_{_{0}}} - {\bf K}} K^{^{1/2}}_{_{+}},
\end{equation}
with ${\bf K} = K_{_{+}}^{^{-1/2}}\, {\cal D}{\cal
N}^{^{\Lambda}}_{_{0}}\, K_{_{+}}^{^{-1/2}},\,\,\bigg|{\bf K}
\bigg| = d $. The formula (\ref{ONEPOLE}) can be presented in a
similar form
\begin{equation}
\label{ONEPOLE1}
 S_{_{0}}(\lambda) = K^{^{-1/2}}_{_{+}} \frac{iI_{_{+}} +
\frac{K^{^{-1/2}}_{_{+}} \phi_{_0} \rangle\,\,\langle
K^{^{-1/2}}_{_{+}}\phi_{_0}}{\lambda - \lambda^{^{\Lambda}}_{_{0}}
}}{iI_{_{+}} - \frac{K^{^{-1/2}}_{_{+}} \phi_{_0}
\rangle\,\,\langle K^{^{-1/2}}_{_{+}}\phi_{_0}}{\lambda -
\lambda^{^{\Lambda}}_{_{0}}}} K^{^{1/2}}_{_{+}} :=
K^{^{-1/2}}_{_{+}} \frac{D_{_{+}} (\lambda)}{ D_{_{-}}(\lambda)}
K^{^{1/2}}_{_{+}}.
\end{equation}
Then
\[
S (\lambda) - S_{_{0}} (\lambda) =
\]
\begin{equation}
\label{deviat}
 K^{^{-1/2}}_{_{+}}\,\,\frac{1}{D_{_{-}}(\lambda)}
\left[ \frac{{\bf K} }{I_{_{+}} - {\bf K}D^{^{-1}}_{_{-}}} \right]
\left( D^{^{-1}}_{_{+}} - D^{^{-1}}_{_{-}}\right)
 D_{_{+}}(\lambda)\,\,K_{_{+}}^{^{1/2}}.
\end{equation}
In above  real neighborhood  the  whole  expression is  small  if
${\bf K}$ is small. Really, denoting  $\psi =
K^{^{-1/2}}_{_{+}}\phi_{_{0}}$ and  denoting by $P_{_{\psi}}$ the
corresponding orthogonal projection,\, $P_{_{\psi}}^{^{\bot}} =
P_{_{+}} - P_{_{\psi}}$, we  have :
\[
D^{^{-1}}_{_{\pm}} = P_{_{\psi}}^{^{\bot}}  + P_{_{\psi}}
\frac{1}{i \pm \frac{\langle K^{^{-1}}_{_{+}}
\phi_{_0},\,\phi_{_0}\rangle}{\lambda -
\lambda^{^{\Lambda}}_{_{0}}}},
\]
\[
D^{^{-1}}_{_{+}} - D^{^{-1}}_{_{-}} = \frac{2\langle
K^{^{-1}}_{_{+}} \phi_{_0},\,\phi_{_0}\rangle ( \lambda -
\lambda^{^{\Lambda}}_{_{0}} )}{\langle K^{^{-1}}_{_{+}}
\phi_{_0},\,\phi_{_0}\rangle^{^2} + ( \lambda -
\lambda^{^{\Lambda}}_{_{0}} )^{^2}} \,\, P_{_{\psi}}
\]
\[
 \left(D^{^{-1}}_{_{+}} - D^{^{-1}}_{_{-}} \right) D_{_{+}} =
\frac{2 \langle K^{^{-1}}_{_{+}} \phi_{_0},\,\phi_{_0}\rangle} {
\langle K^{^{-1}}_{_{+}} \phi_{_0},\,\phi_{_0}\rangle - i (\lambda
- \lambda^{^{\Lambda}}_{_{0}})}P_{_{\psi}}
\]
In particular on real neighborhood $\bigg|D^{^{-1}}_{_{\pm}}\bigg|
\leq 1 ,\,\,\bigg|D^{^{-1}}_{_{+}} - D^{^{-1}}_{_{1}}\bigg| \leq
1,\, \bigg|\left(D^{^{-1}}_{_{+}} - D^{^{-1}}_{_{-}} \right)
D_{_{+}}  \bigg|\leq 2 $ and $ \bigg|\left[I- {\bf K}
D^{^{-1}}_{_{-}} \right]^{^{-1}} \bigg| < \frac{1}{ 1 - d }, $
hence  on real  axis we  have the estimation of the  inner bracket
in (\ref{deviat}) by $\frac{d}{1 - d }$ and the estimation of  the
difference (\ref{deviat}) on the real neighborhood
$\Delta_{_{\varepsilon}}$
\begin{equation}
\label{onpolappr} \bigg|S - S_{_{0}} \bigg| < \frac{d}{1-d}
\sup_{_{\lambda \in \Delta_{_{\varepsilon}}}}
|K^{^{1/2}}_{_{+}}(\lambda)| \sup_{_{\lambda \in
\Delta_{_{\varepsilon}}}} |K^{^{- 1/2}}_{_{+}}(\lambda)|.
\end{equation}
For  small  $q$ and  hence small  $\varepsilon$, the above
estimate  can be  replaced  by  the
\[
\bigg|S - S_{_{0}} \bigg| < \frac{d}{1-d}\, B \,
|K^{^{1/2}}_{_{+}}(\lambda^{^{\Lambda}}_{_{0}})|\,\,|K^{^{-
1/2}}_{_{+}}(\lambda^{^{\Lambda}}_{_{0}})|,
\]
with  some  coefficient  $B$ greater then $1$, for  instance $ B =
3/2 $.
  Similar  estimation  may be  obtained   inside the corresponding
complex neighborhood $U^{^{C}}_{_{\varepsilon}}$  on the  small
circle $C_{_{\varepsilon}}$ centered at  the single  zero
$\lambda_{_{0}} = \lambda_{_{0}}^{^{\Lambda}} + i \langle
K_{_{+}}(\lambda_{_{0}})\phi_{_{0}},\, \phi_{_0}\rangle$ of
$S_{_{0}}$, with radius equal $\bigg| \langle
K_{_{+}}(\lambda_{_0})\phi_{_{0}},\, \phi_{_0}\rangle\bigg|$. Then
we  have  similar  estimate,
\[
 |S - S_{_{0}}| \bigg|_{_{C_{_{\varepsilon}}}} <
\frac{2 d}{1-d} \sup_{_{\lambda \in C_{_{\varepsilon}}}}
|K^{^{1/2}}_{_{+}}(\lambda)| \sup_{_{\lambda \in
C_{_{\varepsilon}}}} |K^{^{-
1/2}}_{_{+}}(\lambda^{^{\Lambda}}_{_{0}})| \approx
\]
\begin{equation}
\label{Rouchet} \frac{2 d}{1-d}\, B\,
|K^{^{1/2}}_{_{+}}(\lambda^{^{\Lambda}}_{_{0}})|
 |K^{^{-
1/2}}_{_{+}}(\lambda^{^{\Lambda}}_{_{0}})|
\end{equation}
where  we  replaced  maximal values of  norms  on the  circle by
values  of  norms  at the  resonance eigenvalue
$\lambda_{_{0}}^{^{\Lambda}}$, assuming that  $\varepsilon$ is
small. If  the  condition
\[
 \frac{2 d}{1-d} B
|K^{^{1/2}}_{_{+}}(\lambda^{^{\Lambda}}_{_{0}})|\,\,
 |K^{^{- 1/2}}_{_{+}}(\lambda^{^{\Lambda}}_{_{0}})| < 1
\]
is  fulfilled, then  there exist, due  to the operator version of
Rouchet theorem,\cite{Gohberg}, an isolated zero of the Scattering
matrix inside the circle $C_{_{\varepsilon}}$. \vskip0.5cm
 Summarizing  results obtained in actual and
previous sections we see that the step-wise structure of
continuous spectrum on the quantum network permits us to reduce
 calculation of  the  Scattering  matrix on the
spectral band  $\Delta_{_{\Lambda}} =
\left[\lambda^{^{\Lambda}}_{_{max}},\,\lambda^{^{\Lambda}}_{_{min}}\right]$
to matching solutions in neighboring  domains  via proper
Dirichlet-to-Neumann map. Hence the large-time asymptotic  of
corresponding non-stationary processes  is defined  by the shape
of eigenfunctions of discrete spectrum of the intermediate
operator. One can use the variational method for the construction
both  the eigenfunctions and eigenvalues of the intermediate
operator below the  threshold $\lambda^{^{\Lambda}}_{_{\min}}$,
and thus obtain the Scattering matrix from the solution of the
variational problem.
\par
The derived estimation of the deviation of the  one-pole
approximation from the Scattering matrix and the  estimation of
positions of corresponding resonances are too crude. The following
``physical" motivation  shows, that possibly contribution from
some non-resonance eigenvalues and from the continuous spectrum in
the previous expression (\ref{Smatrix}) for the Scattering matrix
may be  neglected  under much  wider conditions. This motivation
shows that the one-pole, or possibly few-pole approximations,  may
give a  reasonably complete physical picture of dynamical
processes on the network.
\par
 Assume that $\Lambda$
is the Fermi level in the wires and $T$ is properly scaled
temperature. Then one may obtain an approximate expression for the
Scattering matrix in an essential interval $\left( \Lambda -
T,\,\Lambda + T \right)$ of  energy by substituting  into
(\ref{Smatrix}) properly reduced expression for the  DN-map ${\cal
D}{\cal N}^{^{\Lambda}}_{_T}$ containing singularities only in
that interval. Really, one usually  estimate  the  number of
excited modes in  Fermi-systems  with  discrete  spectrum ,
considering the corresponding  essential interval, because
dynamics of electrons outside the interval is suppressed due to
the Fermi distribution. We just extend  this  principle  to the
case when continuous  spectrum is  present, and implement it via
replacing  of the DN-map of the intermediate operator by the
``essential part" of it :
\[
{\cal D}{\cal N}^{^{\Lambda}} \to {\cal D}{\cal
N}^{^{\Lambda}}_{_T}(\lambda) = \frac{1}{(2 \mu_{_{\Omega}})^{^2}}
    \sum_{_{\left(\Lambda - T\,<\lambda^{\Lambda}_r\,<\Lambda + T \right)}}
\frac{P_{+}\frac{ \partial \varphi_r (x)}{\partial n_x}\rangle
\,\,\langle P_{+}\frac{ \partial \varphi_r (s)}{\partial n_r}}
{\lambda - \lambda^{\Lambda}_r} +
\]
\begin{equation}
\label{reduced} \sum_{V_{m} + \frac{\pi^2 \,l^2}{2 \mu^{\bot}_m
\delta^2_m}> \Lambda}  \frac{1}{(2 \mu_{_{\Omega}})^{^2}}
\int_{_{\left(\Lambda - T\,<\rho\,<\Lambda + T \right)}}
\frac{P_{+} \frac{\partial \varphi_{\rho,l} (x)}{\partial
n_x}\rangle \,\,\langle P_{+} \frac{\partial \varphi_{\rho,l}
(s)}{\partial n_s}} {\lambda - \rho} d\rho.
\end{equation}
Substituting  the the  essential  part  of  the DN-map  into  the
expression  (\ref{Smatrix}) we  obtain the corresponding
``dynamical approximation'' for the Scattering matrix on the
essential interval of energy:
\begin{equation}
\label{Sapprox} S (\lambda) \approx -\frac{{\cal D}{\cal
N}^{^{\Lambda}}_{_T} + i K_+} {{\cal D}{\cal N}^{^{\Lambda}}_{_T}
- i K_+}.
\end{equation}
The reduced  {\it essential}  part ${\cal D}{\cal
N}^{^{\Lambda}}_{_T}$ of the Dirichlet-to-Neumann map of the
intermediate operator  may contain one or several polar terms and
an integral on  an interval of the continuous spectrum. We  will
consider in further text the  case  when  the  essential interval
does  not overlap with  continuous spectrum of the intermediate
operator $\Lambda + T < \lambda^{^{\Lambda}}_{_{min}}$ and neglect
the integral on the continuous spectrum. We call the essential
interval of energy $\left[\Lambda -T,\, \Lambda + T\right]$ {\it
an essential spectral  band }  and denote it  by
$\Delta_{_{\Lambda}}^{^T}$. One may  assume, that  for narrow
wires the continuous  spectrum  does not overlap with essential
spectral band.  The resulting {\it essential} expression
(\ref{Sapprox}) for the Scattering matrix is actually a Scattering
matrix of some solvable model, similar to one constructed in
\cite{BF61,Opening,Zero_range}. In particular we again  obtain the
``one-pole approximation''  (\ref{Sapprox1}) for the Scattering
matrix near the  resonance
\[
S (\lambda) = -\frac{{\cal D}{\cal N}^{^\Lambda} + i K_+} {{\cal
D}{\cal N}^{^\Lambda} - i K_+} \,\,\approx \,\,\displaystyle -
\frac{\frac{P_{+}\frac{ \partial \varphi_{_0} (x)}{\partial
n_x}\rangle \,\,\langle P_{+}\frac{
\partial \varphi_{_0} (s)}{\partial n_s}} {\lambda -
\lambda^{^{\Lambda}}_{_{0}} } + i K_{+} (\lambda)}
{\frac{P_{+}\frac{ \partial \varphi_{_0} (x)}{\partial n_x}\rangle
\,\,\langle P_{+}\frac{ \partial \varphi_{_0} (s)}{\partial n_s}}
{\lambda - \lambda^{\Lambda}_0}-i K_{+} (\lambda)}\,\,,
\]
if only  one  resonance eigenvalue  $\lambda^{^{\Lambda}}_{_{0}}
\approx \Lambda $ of the the  intermediate operator is sitting on
the essential spectral band  $\Delta^{^{T}}_{_{\Lambda}}$.
\vskip0.5cm
 \section{Star-shape model of a Quantum Network}
The  aim of  this  section is:  to construct a  solvable model for
general Quantum network with a  finite number of  semi-infinite
wires. The model network will be constructed  of a single vertex,
where the essential part of the intermediate operator is
substituted by
 some finite-dimensional operator, and several
quasi-one-dimensional quantum wires attached to it, emulating the
open channels. We will choose the parameters of the model such
that the corresponding Scattering matrix  will coincide with the
essential Scattering matrix  of the Quantum Network
(\ref{Sapprox}). Similar construction for the simplest Quantum
Network - a three-terminal Quantum Switch - was developed in
preprint \cite{MPP03}, see Fig.1 below. We will  use in this
section notations similar to  ones introduced in section 2
\[
{ U}_{_{_{\Omega}}}= \sum_t \oplus \,\,u_{_{\Omega_t}},\,\, {
U}^{^{\omega}}= \sum_{_{t,m}} \oplus \,\,{
u}^{^{\omega}}_{_{t,m}},
\]
but  we  assume  now  that  vector  functions
${U}_{_{_{\Omega}}},\,{U}^{^{\omega}}$ present  elements of
functional spaces not on  the  bottom sections  of channels, but
 on the  vertex  domains and  wires respectively. With these
notations we can present the basic Schr\"{o}dinger equation on the
vertex domains and  on the wires in symbolic form
\[
 -\frac{1}{\mu_{_{\Omega}}} \bigtriangleup {U}_{_{\Omega}}
 + V_{_{\Omega}} {U}_{_{\Omega}} = E\, {U}_{_{\Omega}}
\]
\begin{equation}
\label{Symbeq} -\frac{1}{\mu^{^{\parallel}}}\frac{\partial^{^2}
{U}^{^{\omega}}}{\partial x^{^2}}  +
\frac{1}{\mu^{^{\bot}}}\frac{\partial^{^2}
{U}^{^{\omega}}}{\partial y^{^2}} + V^{^{\omega}}\,
{U}^{^{\omega}} = E \, {U}^{^{\omega}},
\end{equation}
with a  bounded measurable potentials  $V_{_{\Omega}} =
\left\{V_{_{t}} \right\}$ on the wells and constant potentials
$V^{^{\omega}} = \left\{V_{_{m}} \right\}$ on the  wires. Consider
the  the  entrance eigen-vectors $e^{^m}_{_l}$ of  cross-section
 and  corresponding  orthogonal projections  $p^{^m}_{_l}$
 in  spaces of square-integrable functions  on
  bottom sections  of the  wires $\omega^{^m}$,
see section 2, and form  the  entrance  subspaces  $E_{_{\pm}}$
for open and closed  channels with corresponding projections
\[
P_{_{\pm}} = \sum_{_{\pm}}\oplus p^{^m}_{_l},\,\,\mbox{dim}
E_{_{+}} = N
\]
with  summation over  the  thresholds below and  above the  Fermi
level  $\Lambda$ respectively. For  instance: we  take for
$E_{_{+}}$ the sum of all $p^{^m}_{_l}$ with indices satisfying
the condition $\frac{1}{2 \mu_{_{m}^{^{\bot}}}}\, \frac{\pi^{^2}
l^{^{2}}}{\delta_{_{m}}^{^{2}}} + V_{_{m}} < \Lambda $ for $
E_{_{+}} $. For values  of  the  spectral parameter on  the
minimal spectral band $\left[ \lambda_{_{min}}^{^{L}},\,
\lambda_{_{max}}^{^{L}}\right]$ containing the  Fermi level
$\Lambda$, we introduce an  operator-valued Wave-number ${\bf
P}_{_{+}}$ for open channels as  a positive  square root  from
\[
{\bf P}^{^2}_{_{+}} = \sum_{_{\left\{ +\right\}}} \left\{\lambda -
\frac{1}{2 \mu_{_{m}^{^{\bot}}}}\, \frac{\pi^{^2}
l^{^{2}}}{\delta_{_{m}}^{^{2}}} - V_{_{m}}\right\} p^{^m}_{_l} :=
\lambda - {\bf V}^{^{\omega}},
\]
with  summation $\left\{ +\right\}$ over all  open  channels, and
the positive  decrement  $P_{_{-}}$ for  closed  channels:
\[
{\bf P}^{^2}_{_{-}} = \sum_{_{\left\{ -\right\}}} \left\{
\frac{1}{2 \mu_{_{m}^{^{\bot}}}}\, \frac{\pi^{^2}
l^{^{2}}}{\delta_{_{m}}^{^{2}}} + V_{_{m}} - \lambda \right\}
p^{^m}_{_l} =  {\bf V}^{^{\omega}} - \lambda.
\]
The  introduced operators are connected  with operators
$K_{_{\pm}}$ from the previous  section
\[
K_{_{\pm}} = \sqrt{2\mu^{^{\parallel}}}{\bf P}_{_{\pm}}.
\]
Corresponding  exponential  solutions  of the  Schr\"{o}dinger
equation in the wires are:
\[
e^{^{\pm i\,K_{_{+}}\,x }}\,\,e,\, \,\mbox{with}\,\, e \in
E_{_{+}}\,,\,\mbox{and}\,\, e^{^{\pm K_{_{-}}\,x}},\,\,e\,\,
\mbox{with}\,\, e \in E_{_{-}}.
\]
Depending on  architecture of the  Quantum Network one  may
construct the  intermediate operator either  blocking  open
channels  in  {\it all  wires}, as  suggested at the end of
section 2, or blocking  open channels  only in {\it semi-infinite
wires}. When constructing  the  solvable  model of the  network we
also have  a  choice: either  we  present  the whole  network as a
single  star-shaped  graph with a single  resonance vertex, or as
a joining of  star-shaped  graphs with resonance vertices,
connected  by finite  quantum wires. In second case  the
parameters of the finite wires enter the  model as explicit
parameters. We guess that the second representation is  more
convenient for detailed engineering analysis of networks and
especially for optimization of their transport properties.
Nevertheless  we choose  now the former construction,  firstly: as
more elementary and, secondly: just to  include in our
construction non-trivial ( bent of periodic) internal finite
wires. When analysis of the first construction is done, we can, at
least in principle, develop analysis for the second one, with
straight connecting wires, in algebraic way,  based on star-shaped
elements. Note that importance  of  star-shaped elements  was
noticed long ago and remains  motivation of numerous  recent
papers, see for instance \cite{HarmerTh,Exner03}. We  prove  here
that the star- shaped  graph  with a ``resonance vertex'', see our
construction below, is an universal  object in spectral  theory of
Quantum Graphs .
\par
When chopping  off  the  open channels  in  semi-infinite wires
only, we  obtain  a  self-adjoint  intermediate  operator
$L_{_\Lambda}$  and  construct the  corresponding  DN-map ${\cal
D}{\cal N }^{^{\Lambda}}$  and  the  approximation  of the
Scattering matrix  on the  essential spectral band,
(\ref{Sapprox}). The  corresponding  Scattering Ansatz
(\ref{Sansatz_open}) in semi-infinite open channels satisfies the
equation
\begin{equation}
\label{nonpert}
 -\frac{1}{\mu^{^{\bot}}}\, \frac{d^{^2} {U}^{^{\omega}}}{d x^{^2}}
  + {\bf V}^{^{\omega}} {U}^{^{\omega}} = \lambda \,\, {U}^{^{\omega}}.
\end{equation}
 We  choose the outer (``trivial") the part  of  the  model -
 in  open  channels - in form of  original   matrix
 Schr\"{o}dinger operator
\[
l \,\,\,{U}^{^{\omega}} =  -\frac{1}{\mu^{^{\bot}}}\, \frac{d^{^2}
{U}^{^{\omega}}}{d x^{^2}} + {\bf V}^{^{\omega}} {U}^{^{\omega}}
\]
in $L_{_{2}}(E,\,[0,\infty))$ and  we  will  use  the  same
Scattering Ansatz (\ref{Sansatz_open}) in  open wires  as for the
original  problem. But the construction of the vertex part of the
model will be done with a major change of the original
intermediate operator.
\par
 Based on the above
physical motivation  we  assume  that  only  a  finite number
$N_{_{T}}$ of eigenvalues of the  intermediate  operator sits on
the essential spectral band and  no change of multiplicity  of the
continuous  spectrum  occurs. Then  we  will substitute the
intermediate operator on the  Network  by  some
 finite-dimensional positive  hermitian  matrix $A$ acting in
an abstract space $E_{_{A}}$, dim$\, E_{_{A}} = \,N_{_T}$. The
eigenvalues and normalized eigenvectors of the operator $A$ will
be  denoted by $ k_{_s}^{^2},\, e_{_s}$ correspondingly, $\, s=
1,2,\dots N_{_{T}} $. The eigenvalues and the boundary parameters
of the model $\,\beta$, see below (\ref{one}), will be defined
later based on comparison of the Scattering matrix of the model
with the essential Scattering matrix. The  resulting  operator
${\bf A}_{_\beta}$  we  obtain  is  a  simplest  star-shaped model
of  the  Quantum network  with the  compact part  substituted  by
a  sophisticate  Quantum Dot. In our construction we follow the
receipt of construction of zero-range potentials with inner
structure described  in \cite{Extensions,Kurasov}, see also
references therein.
\par
\begin{figure}[h]
\begin{center}
\includegraphics[height=5cm,width=6cm]{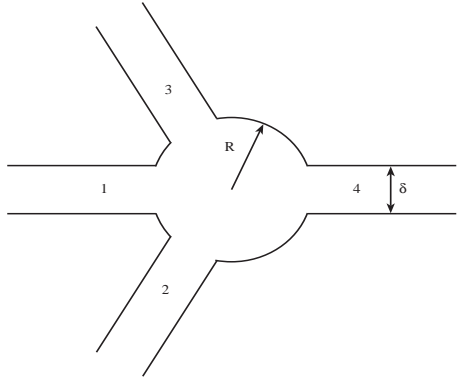}
\caption{Three-terminal Quantum Switch}
\end{center}
\end{figure}
\noindent The  corresponding  model for  the  three-terminal
Quantum Switch is used in  \cite{PRB} for estimation  of  the
parameter regime of the  switch.
\par
 Restrict both $l$ and $A$ to symmetric operators
onto the corresponding domains: $l\to l_{_0} = l
\bigg|_{_{D^{^{l}}_{_{0}}}}$ defined on functions vanishing near
near $x = 0$; the operator is  restricted $A\to A_{_{0}} = A
\bigg|_{_{D^{^{A}}_{_{0}}}}$ such that  given subspace  $N \in
E_{_A} $  would  play  the  role  of  deficiency subspace at the
spectral point  $i: N = N_{_i}$, dim $N = d $. Then dual
deficiency subspace is ${ N}_{-i} = \frac{A + i I}{A - i I} N$.
Define  the domain of the  restricted operator $A_{_{0}}$ as
$D^{^{A}}_{_{0}} = \frac{I}{A - iI}\,\,[E_{_{A}} \ominus
N_{_{i}}]$ of $A_{_{0}}$. It is  not dense, since $A$ is bounded.
Nevertheless, under condition  that the deficiency subspaces
$N_{_{\pm}} = N_{_{\pm i}}$ do not have  common elements, see for
instance \cite{Krasn,KP00}, the extension procedure  for the
orthogonal sum $l_{_0}\oplus A_{_{0}}$ can be developed with use
of the symplectic formalism, see for instance \cite{Zero_range}.
The  case dim$E_{_{A}} = 1 $ is  not  formally  covered by  the
above condition, but  the  corresponding extension also  was
constructed in  \cite{Shirokov79,Shirokov80} and  the  relevant
formulae for the scattering  matrix and scattered waves remain
true.The extension of the restricted operator ${\bf A}_{_{0}}$ is
reduced to selection of a Lagrangian plane of the sum of the
corresponding boundary forms, see for instance
\cite{Extensions,KP00}, and below, see \ref{one}, where the
Lagrangian plane is  chose  based  on proper  boundary condition.
The  same  description of  extensions  remains valid in case  dim$
E_{_{A}} = 1$.  We  will use  this fact below, see {\ref{bcond0}}.
\par
The  boundary form  of  the  differential operator
$l_{_{0}}^{^{+}}$ is obtained via  standard integration by parts:
\begin{equation}
\label{boundform} {\cal J}_{_l} (U,\, V) =  \langle
l_{_{0}}^{^{+}} U,\,V\rangle - \langle U ,\, l_{_{0}}^{^{+}}
V\rangle = \langle U'(0),\, V (0)\rangle  - \langle U(0),\,
V'(0)\rangle,
\end{equation}
where  $U (0),\,V(0) \in E_{_{+}}$ and  the  derivatives are taken
in positive direction.
\par
We assume that the Hermitian operator $A : E_{_{A}} \to E_{_{A}}$,
dim $E_{_{A}} = N_{_{T}}$, is defined by its spectral
decomposition (on discrete spectrum)
\[
A = \sum_{_s}^{^{N_{_{T}}}} k^{^2}_{_s} P_{_{s}},
\]
where  $k^{^2}_{_s}$ are  positive  eigenvalues  and  $P_{_s}$ are
the corresponding spectral  projections, $P_{_{s}}=
e_{_{_s}}\rangle \langle e_{_{s}}$. Choose an ortho-normal basis
in ${N}_i$ $\left\{f_l\right\} = \left\{f_{l,i}\right\},\,l=
1,2,\dots ,d,\, d =\dim {{ N}_i} \leq N_{_{T}} $, as a set  of
deficiency vectors  of the  restricted  operator  $A_0$. Then
vectors $\frac{A + i I}{A - i I} f_l$ form an ortho-normal basis
in the dual deficiency subspace $N_{_{-i}}$. We  assume  that the
subspaces ${N}_i$ and ${N}_{-i}$ do not overlap, hence their
(direct) sum ${\cal N}={N}_i {+}{ N}_{-i}$,- the {\it defect} of
the operator $A_{_0}$, - has the dimension $2 d = 2$. Under this
assumption the deficiency  index of the restricted operator $A_0$
is equal to $(d, d)$.  The  defect contains elements from the
domain $D_0$ of $A$,  so the restricted operator $A_0 =
A\big|_{D_0}$ is not densely defined, and hence the adjoint
operator does not exist. However, under the above condition one
may  use the formal adjoint $A_0^+$ defined on defect:
$$u =  \sum_{l=1}^d [x_{l,i}\, e_{l,i} + x_{l,-i} \, e_{l,-i}]
\,\in \, {\cal N},
  $$
 by the von-Neumann formula, see for  instance \cite{Glazman}, $$A_0^+ u =
 \sum_{l=1}^d [ -i\,\, x_{l,i}\, f_{l,i} + i\,\, x_{l,-i}\, f_{l,-i}].$$
 We  will  use the formal adjoint during  the construction of  extensions  (see
  \cite{Krasn,KP00})  since, for closed operators  with the finite deficiency
  indices, the construction of  the  extension is actually developed  in  the defect.
  In order  to use the symplectic version of the operator-extension
 techniques we introduce in  the defect ${\cal N}$
 a  new basis $w_{\pm,l}$, on which the formal adjoint $A_0^+$  is
 correctly defined due  to  the  above  assumption of non-overlapping:
 \[
  w_{_{l,+}}= \frac{f_{_{l,i}} + f_{_{l,-i}}}{2} = \frac{A}{A-iI} f_{_l},
  \,\,w_{_{l,-}} = \frac{f_{_{l,i}} - f_{_{l,-i}}}{2 i}= - \frac{I}{A-iI}
  f_{_l},
  \]
   \[
  A_0^+ w_{_{l,+}} = w_{_{l,-}},\,\, A_0^+ w_{_{l,-}} = - w_{_{l,+}}.
  \]

  The following  formula of integration by parts with abstract operators
  was proved in (\cite{Zero_range,Extensions}):

  \begin{lemma}
  {
       Consider  elements from
       the  domain of  the  (formal) adjoint  operator $A_0^+ :$
       $$
       u = u_0 +\sum_l[ \xi^{^u}_{_{l,+}} w_{_{l,+}} + \xi^{^u}_{_{l,-}} w_{_{l,-}}] =
       u_0 + \frac{A}{A-iI} \vec{\xi}^{^u}_{_{+}} - \frac{1}{A-iI}
       \vec{\xi}^{^u}_{_{-}},
       $$
       $$
        v = v_0 + \frac{A}{A-iI} \vec{\xi}^{^v}_{_{+}} -
        \frac{1}{A-iI} \vec{\xi}^{^v}_{_{-}}
        $$
with  symplectic  coordinates  $
\vec{\xi}^{u}_{\pm},\,\vec{\xi}^{v}_{\pm}$:
       \[
 \vec{\xi}^{^u}_{_{\pm}} = \sum_l \xi^{^u}_{_{l,\pm}} f_{_{l,i}}\in {N},\,\,
  \vec{\xi}^{^v}_{_{\pm}} = \sum_l \xi^{^v}_{_{l,\pm}} f_{_{l,i}}\in {N}.
       \]
       Then,  the
       boundary  form  of the  formal adjoint operator  is  equal to
               \begin{equation}
        \label{b_form}
       {\cal J}_{_{A}} (u,v) = \langle A_0^+ u,v \rangle - \langle  u, A_0^+ v
       \rangle = \langle \vec{\xi}^{^u}_{_{+}}, \vec{\xi}^{^v}_{_{-}}\rangle_{_N} -
       \langle \vec{\xi}^{^u}_{_{-}}, \vec{\xi}^{^v}_{_{+}}
       \rangle_{_N}.
        \end{equation}
}
        \end{lemma}
       The fundamental Krein formula \cite{K,MN}
       for Generalized Resolvents of
      Symmetric Operators  can  be easily derived from
       the  next statement proved  in\cite{Extensions}.
       It helps to solve the  non-homogeneous
       equations  with the adjoint  operator  $(A^{^+}_{_0} -\lambda I)u = f$.
        In our situation  similar to  \cite{Ad_Pav}
       it is used in course of calculation of the scattering
       matrix , see  below (\ref{model}).
       \begin{lemma} {The symplectic coordinates
       $\vec{\xi}^{^u}_{_\pm}\in N $ of
        the  vector-valued  function
       \begin{equation}
       \label{incomp}
        u =  \frac{A +iI}{A - \lambda I}\,\,\, \vec{\xi}^{^u}_{_{+}} :=
        u_0 + \frac{A}{A-iI} \vec{\xi}^{^u}_{_{+}}
        - \frac{1}{A-iI} \vec{\xi}^{^u}_{_{-}},
        \end{equation}
        which  satisfies the adjoint  equation  $ [A_0^+ - \lambda I] u =
        0$, are  connected  by  the  formula
       \begin{equation}
       \label{krein}
       \vec{\xi}^{^u}_{_{-}} = - P_{N}\frac{I + \lambda A}{A -
       \lambda} \vec{\xi}^{^u}_{_{+}}
       \end{equation}
       }
       \end{lemma}
       The operator-function
       $$ {\cal M}=  P_{ N} \frac{I + \lambda A}{A - \lambda I}
P_{N}
       : {N} \to { N}, $$
        with a  positive  imaginary  part  in the upper half-plane
       $\Im \lambda > 0$ coincides actually with  ``$Q$-function",
       introduced by M. Krein (\cite{K}). It serves an abstract analog
        of the celebrated function used  by  H. Weyl in  course  of
       construction of  self-adjoint extensions of the  second  order
       differential operator  of  Schr\"{o}dinger  type, see, for  instance
        \cite{Titchmarsh}, where the original construction  by  H.Weyl is
       presented. The operator-function ${\cal M}$ exists almost
       everywhere on real axis $\lambda$
       and  has a   finite  number
       of  simple  poles  sitting at  the  eigenvalues  $k_{_l}^{^2}$ of $A$.
       This function  plays  an  important role in description of spectral
       properties of self-adjoint extensions of  symmetric operators, see
       \cite{K,Zero_range,Gorbachuk}.
       \par
       The boundary form ${\bf J} ({\bf U},{\bf V}) $ of the orthogonal
       sum ${\bf A}_0^+ = A_0^+ \oplus {l}_0^+ $ of the restricted
       matrix and the differential operator on elements $(U,u) = {\bf
       U}$ on the orthogonal sum  of the corresponding spaces  is
       calculated as a sum of the forms (\ref{boundform},\ref{b_form}). The
       self-adjoint extensions of  the corresponding operator  ${\bf A}_0
       = A_0 \oplus {l}_0$ are obtained  as restrictions of  the
       adjoint operator  ${\bf A}_0^+= A_0^+ \oplus {l}_0^+$ onto
       Lagrangian planes of the form ${\bf J} ({\bf U},{\bf V})$. These
       planes may be defined by the boundary conditions  connecting the
       symplectic coordinates $ U'(0),\,U (0) ,\,\, \vec{\xi}^u_{+},\,
        \vec{\xi}^u_{-} $ of components of the corresponding  elements in the
       deficiency subspaces. For instance, one may select a
       $d$-dimensional operator
       $\beta :E \to {\cal E} $ and define the Lagrangian plane
       ${\bf L}_{\beta}$ by the boundary condition
       \begin{equation}
       \label{one}
        \left(
       \begin{array}{c}
       U'(0)\\ -\vec{\xi}_{+}
       \end{array}
       \right)
       =
        \left(
       \begin{array}{cc}
       \beta_{_{00}} & \beta_{_{01}}\\ \beta_{_{01}}^{^{+}} & 0
       \end{array}
       \right)
        \left(
       \begin{array}{c}
       U(0) \\ -\vec{\xi}_{-}
       \end{array}
       \right).
       \end{equation}
This  condition defines a  self-adjoint extension ${\bf
A}_{\beta}$  of the restricted operators $A_0\oplus l_0$ in
$L_2(R,{E_{_{+}}}) \oplus E_{_{A}} $. The absolutely  continuous
of the operator ${\bf A}_{\beta}$ coincides  with  the spectrum of
the exterior part  of the model, hence - with the  spectrum  of
the trivial component $l_{_{\Lambda}}$ of the split  operator
${\cal L}_{_{\Lambda}}$ . The  corresponding
 eigenfunctions of ${\bf A}_{_{\beta}}$
 on the  minimal spectral band
 $\lambda_{_{\max}}^{^{\Lambda}}\lambda_{_{min}}^{^{\Lambda}}$
 containing the  Fermi-level  can be  found
 via  substitution  into  above  boundary  condition the column
 combined  of  Anzatzes  (\ref{Sansatz_open}) and  (\ref{incomp})
 \[
\Psi = \left(
\begin{array} {c}
e^{-i K_+ x} \nu  + e^{ iK_{+}x} {\bf S} \nu\\
\frac{A + iI}{A - \lambda I}\vec{\xi}^{u}_{+}
\end{array}\right).
 \]
It  gives the linear equation for  Scattering matrix:
\[
\left(
\begin{array}{c}
-iK_{_{+}} \left( \nu - S \nu\right)\\
-\vec{\xi}_{_+}
\end{array}
\right) = \left(
\begin{array}{cc}
\beta_{_{00}}& \beta_{_{01}}\\
\beta_{_{00}}^{^{+}}& 0
\end{array}
\right) \,\, \left(
\begin{array}{c}
\nu + S \nu\\
{\cal M}\vec{\xi}_{_+}
 \end{array}
 \right).
\]
Solving  this  equation we can obtain  the  Scattered  waves and
the formula  we  needed:
\begin{lemma}
{The Scattering matrix  for  the  constructed  model is an
analytic function of the  spectral parameter $\lambda$:
\begin{equation}
\label{model}
 {\bf S}(\lambda) = \frac{iK_{_{+}} +\left[\beta_{_{00}} -
\beta_{_{01}} {\cal M} \beta_{_{10}} \right]}{iK_{_{+}}
-\left[\beta_{_{00}} - \beta_{_{01}} {\cal M} \beta_{_{10}}
\right]},
\end{equation}
with denominator of the  fraction preceding the  numerator. }
\end{lemma}
Now  we  will choose the parameter $\beta$ such that the
operator-function $\left[\beta_{_{00}} - \beta_{_{01}} {\cal M}
\beta_{_{10}} \right]$ acting  in $E_{_{+}}$ coincides with the
essential component  ${\cal D}{\cal N}^{^{\Lambda}}_{_{T}}$ of the
DN-map of the intermediate operator. Really, denote by $Q_{_{s}}$
the spectral projection  onto the eigen-space of $A$ corresponding
to the eigenvalue $k^{^2}_{_s}$ framed by projections onto the
deficiency subspace $N_{_{i}} = N$
\[
Q_{_{s}} = P_{_{i}} e_{_{s}}\rangle \langle  P_{_{i}} e_{_{s}}.
\]
Then the  above expression takes the form :
\[
\left[ \beta_{_{00}} - \beta_{_{01}} {\cal M} \beta_{_{10}}
\right]=
\]
\begin{equation}
\label{KreinQ} \left[\beta_{_{00}} + \sum_{_{s=1}}^{^{N_{_{T}}}}
k_{_s}^{^2} \beta_{_{01}} Q_{_{s}} \beta_{_{10}}\right] +
\sum_{_s} \frac{1 + k_{_{s}}^{^4}}{k_{_{s}}^{^2} - \lambda
}\beta_{_{01}} Q_{_{s}}\beta_{_{10}}.
\end{equation}
We will define the boundary parameters
$\beta_{_{10}},\,\,\beta_{_{01}}  = \beta_{_{10}}^{^ +}$ later,
but once they are defined, we choose $\beta_{_{00}}$ such that the
first summand in the  right side of (\ref{KreinQ}) vanishes:
$\beta_{_{00}} + \sum_{_{s}} k_{_s}^{^2}\beta_{_{01}} Q_{_{s}}
\beta_{_{10}} = 0$. Then  the Scattering matrix  take the form:
\begin{equation}
\label{ModelSmatr} {\bf S}(k) = \frac{i K_{_{+}} -
\sum_{_{s=1}}^{^{N_{_{T}}} \frac{1 + k_{_{s}}^{^4}}{k_{_{s}}^{^2}
-\lambda} \beta_{_{01}} Q_{_{s}}\beta_{_{10}}}}{i K_{_{+}} +
\sum_{_s} \frac{1 + k_{_{s}}^{^4}}{k_{_{s}}^{^2} - \lambda}
\beta_{_{01}} Q_{_{s}}\beta_{_{10}}},
\end{equation},
which  coincides  with  the  essential  Scattering matrix  if  and
only  if  the  corresponding   Krein -function
\begin{equation}
\label{Kreinfunk}
 \sum_{_{s=1}}^{^{N_{_{T}}}} \frac{1 +
k_{_{s}}^{^4}}{k_{_{s}}^{^2} -\lambda} \beta_{_{01}}
Q_{_{s}}\beta_{_{10}}
\end{equation}
coincides  with  with the essential  part ${\cal D}{\cal
N}^{^{\Lambda}}_{_{T}}$ of the $DN$-map  of the intermediate
operator - the finite sum of polar terms
\begin{equation}
\label{polarterm}
\sum_{_{s=1}}^{^{N_{_{T}}}}\frac{P_{_{+}}\frac{\partial
\varphi_{_{s}}}{\partial n}\rangle \langle P_{_{+}}\frac{\partial
\varphi_{_{s}}}{\partial n}}{\lambda_{_s} - \lambda}.
\end{equation}
Summarizing above results we  obtain the  following conditional
statement  statement for  the  extension constructed  based on the
boundary condition (\ref{one}) in case  when  $N_{_{i}} \cap
N_{_{-i}} = 0$ or  dim $E_{_{A}} = 1$:
\begin{theorem}
{The  constructed  operator   ${\bf A}_{_{\beta}}$ is  a  solvable
model  of  the  Quantum network on the essential interval of
energy  if  and  only  if  the dimension of  the  space $E_{_{A}}$
coincides  with  the  number  of  eigenvalues  on the  essential
interval of  energy, the eigenvalues $k^{^2}_{_{s}}$  of the
operator  $A = \sum_{_{s=1}}^{^{N_{_{T}}}} k^{^{2}}_{_{s}}
e_{_s}\rangle \, \langle e_{_s}$, with some ortho-normal basis
$e_{_{s}}$, coincide with eigenvalues of the intermediate operator
on essential interval of energy  and there
 exist a  subspace $N_{_{i}} = N$ and operator
 $\beta_{_{01}}: N_{_{i}} \to E_{_{+}}$   such  that  for that
 ortho-normal basis $\left\{ e_{_s}\right\}_{_{s = 1}}^{^{N_{_{T}}}}$
 in $E_{_{A}}$
\begin{equation}
\label{}
 P_{_{+}}\frac{\partial
\varphi_{_{s}}}{\partial n} =  \beta_{_{01}} P_{_{i}} e_{_s}
\end{equation}
 for  all  $s,\, s= 1,2,3,\dots ,N_{_{T}}$.
 }
\end{theorem}
 In case when only one resonance
eigen-value $k^{^2}_{_0}$ of the intermediate operator sits  on
the essential spectral band the
 obtained  model Scattering matrix
\begin{equation}
\label{one_pole} {\bf S}(p) = \frac{i K_{_{+}} - \frac{1 +
k_{_{0}}^{^4}}{k_{_{0}}^{^2} - \lambda }\beta_{_{01}}
Q_{_{0}}\beta_{_{10}}}{i K_{_{+}} +  \frac{1 +
k_{_{0}}^{^4}}{k_{_{0}}^{^2} - \lambda} \beta_{_{01}}
Q_{_{0}}\beta_{_{10}}}
\end{equation}
is  a one-pole approximation of  the Scattering matrix  of the
network. The  condition of the  above  theorem  are obviously
fulfilled for one-pole  approximation, when
$P_{_{+}}\frac{\partial \varphi_{_{0}}}{\partial n} \neq  0$,  $d
= 1,\,N_{_{T}} = 1$, and $\beta_{_0}$ is  one-dimensional operator
mapping the one-dimensional subspace $N_{_{i}}$ onto  the
resonance entrance subspace  in $E_{_+}$ spanned  by $
P_{_{+}}\frac{\partial \varphi_{_{0}}}{\partial n}$. For thin or
shrinking  networks  one can estimate the deviation of the
one-pole approximation from  the  exact Scattering matrix on the
network based  on  material of section 3. It will be done
elsewhere.
\par
Note  that there  exist  an  intrinsic  connection between the
matrix model of  the  network  we  propose  and  method  of
derivation  of  the  asymptotic of  different  series  of
eigenfunctions of  Laplace operator  on  a  composite  domain with
shrinking  channels : both  are  actually  based  on  a  shape  of
 eigenfunctions in limiting  domain. We  hope  to  develop  a
 comparative analysis  of both problems  in  a  forthcoming
 publication.
\par
The above one-pole approximation has  also  another important role
in perturbation theory: as  a  sort of a ``jump-start" in analytic
perturbation  procedure. This role  may be clarified by the
following observation. Consider a simplest  network with only one
semi-infinite wire  and  the Fermi level sitting on the first
spectral band  multiplicity $1$. Then the component of the
corresponding solvable model in the open channel is  presented by
 the  Schr\"{o}dinger equation
 \[
- u'' = \lambda u: = p^{^2} u,\, \, 0< x < \infty,
 \]
with one-dimensional  entrance  subspace  $E_{_{+}}$. Construct
the solvable  model ${\bf A}_{_{\beta}}$ of the  network assuming
that the ``inner Hamiltonian" $A$ in course of construction of the
solvable model is restricted to a symmetric  operator with
one-dimensional deficiency  subspace $N_{_{i}} := N $ spanned
 by the vector $e$, and $\beta_{_{01}}:= \beta$ is  the map of $N$
 onto  $E_{_{+}}$. Then the model Scattering matrix is scalar :
\begin{equation}
\label{scalarsmatr} {\bf S}_{\beta}(p) = \frac{ip -
 \beta^{^2} \sum_{_s}
\frac{1 + k_{_{s}}^{^4}}{k_{_{s}}^{^2} - p^{^2}} Q_{_{s}}}{ip +
\beta^{^2} \sum_{_s} \frac{1 + k_{_{s}}^{^4}}{k_{_{s}}^{^2} -
p^{^2}} Q_{_{s}}}
\end{equation}
with  $Q_{_s} = |\langle e,e_{_s}\rangle|^{^2}$. Zeroes  $k_{_{s}}
(\beta)$ of the Scattering matrix (\ref{scalarsmatr}) - resonances
- sit in upper  half-plane and approach  the  points $ \pm
k_{_s},\, k_{_{-s}} = - k_{_s}$, when $\beta \to 0$. They can be
found approximately in form of branching continuous fractions, see
\cite{Voevodin} as  solutions  of a  chain of algebraic equations.
For instance  the  resonance $p_{_0} = k_{_{0}}(\beta)$ created
from the eigenvalue  $k^{^2}_{_0}$ of the operator $A$ ( more
precise : from the point  $+ k_{_0}$ ) can be obtained as  a
solution  $p = k_{_{0}}(\beta)$ of the  equation
\begin{equation}
\label{disp}
 p = k_{_{0}} - \frac{\beta^{^2} (1+ k^{^4}_{_{0}})
Q_{_0} }{(p + k_{_{0}} )\left( ip - \beta^{^2} \sum_{_{s\neq 0}}
\frac{1 + k_{_{s}}^{^2}}{k_{_{s}}^{^2} - p^{^2}} Q_{_{s}}\right)}.
\end{equation}
Another  resonance  created at the  point  $- k_{_0}$
 from the  eigenvalue  $k^{^2}_{_0}$ sits at the  symmetric point
$ -\bar{k}_{_{0}}(\beta)$ with respect to  the  imaginary axis.
Other resonances  $k_{_s} (\beta)$  can be  found   from similar
equations  with  $\pm k_{_{0}}$ substituted  by  $\pm k_{_{s}}$.
All  functions  $k_{_{s}} (\beta) $ are  analytic functions  of
$\beta,\, k$ in small neighborhoods  of  $(0,\, \pm k_s)$. They
sit in the upper  half-plane symmetrically with respect to the
imaginary axis  $k_{_{- s}} (\beta) = - \bar{k}_{_{s}} (\beta) $.
The Scattering  matrix (\ref{scalarsmatr}) is unitary on the  real
axis  $k$ and  has poles at  the  complex-conjugate  points
$\bar{k}_{_{s}} (\beta)$ in the  lower half-plane , hence  it  is
presented  by  the finite Blashke  product  which tends  to  $1$
when  $|k|\to \infty$:
\begin{equation}
\label{SBlashke} {\bf S}^{^{\beta}} (p) =  \prod_{s}\frac{p -
k_{_s}(\beta)}{p - \bar{k}_{_s}(\beta)}.
\end{equation}
The  outer  component of the  scattered  wave  is  presented as
\[
\Psi^{^{\beta}}_{_{0}} = e^{ - ipx}  +  {\bf S}^{^{\beta}}(k) e^{-
ipx},
\]
and  fulfills  proper  boundary  condition  at the  place  of
contact with the model quantum dot.
\par
Though the resonances  depend analytically  of $(\beta,\,p\,)$,
neither  Scattering matrix  (\ref{scalarsmatr}) nor the  Scattered
wave  depend  analytically  of  $ \left(\beta, p\right)$
 on the  product  $(U_{{\beta}}, C_{_p})$ of a  small
neighborhood  of  the  origin  in $\beta$ - plane  and  the  whole
$p$-plane. The  analyticity is lost due to presence of points $\pm
k_{_{s}}$ where the  resonances  are  created at the  origin of
the $\beta$-plane. But it is present for  ``separated"  factors of
the Scattering Matrix. Consider for small $\beta$ two non
intersecting subsets $K_{_0}$ and $K^{^0}$ of the  set   $K =
\left\{ k_{s}\right\}$
 and  the  corresponding factors  ${\bf S}^{^{\beta}}_{_0},\,
{\bf S}_{_{\beta}}^{^0}$  of the  Scattering matrix  with
resonances created  in  $K_{_0},\,K^{^0}$  respectively:
\[
{\bf S}^{^{\beta}}_{_{0}} (p) =  \prod_{_{k_{_s}\in K_{_0}}}
\frac{p - k_{_s}}{p - \bar{k}_{_s}},\,\,\,\,\, {\bf
S}_{_{\beta}}^{^{0}} (p) =  \prod_{_{k_{_s}\in K^{^0}}} \frac{p -
k_{_s}}{p - \bar{k}_{_s}}.
\]
Here  the  first factor  is  an  analytic function of  $(\beta,\,
p )$ when $\beta$ is  small and $p$ sits near   $K^{^0}$, and ,
vice versa, the second factor is analytic for small $\beta$ and
$p$ sits in a small neighborhood of $K_{_0}$. In particular the
first factor can be expand into Taylor series in $(\beta,\, p)$
for small $\beta \in U_{{\beta}}$ if $p$ sits in some  small
neighborhood $U^{^0}$ of $K^{^0}$, and  the second factor is
presented  by the Taylor series in
 $(\beta,\, p)$ if  $\beta \in U_{{\beta}}$  and   $p$
sits  in a  small  neighborhood   $U_{_{0}}$  of $K_{_0}$. On the
other hand, the first factor is not an analytic function  of
$(\beta,\,p)$  if  $\beta \in  U_{_{\beta}}$
 and $p$ is near to $K_{_0}$. This
 observation may be interpreted  as a  fact  of  the
perturbation theory  the  following  way.
\par
Assume that $k_{_0} > 0$ and  consider the representation of the
Scattering  matrix (\ref{SBlashke}) in form of  product of two
factors : the  product containing  the  resonances $k_{_0}
(\beta),\, -\bar{k}_{_0}(\beta)$  created  from the eigenvalue
$k^{^2}_{_0}$
\[
{\bf S}^{^{\beta}}_{_{0}} (p) = \frac{[p - k_{_0}(\beta)]\, [p +
\bar{k}_{_0}(\beta)]}{[p - \bar{k}_{_0}(\beta)]\,[p +
k_{_0}(\beta)]},
\]
and the  complementary  factor ${\bf S}^{^{\beta}}_{_{0}} $
\[
{\bf S}_{_{\beta}}^{^{0}} (p) = \prod_{s\neq 0}\frac{p -
k_{_s}(\beta)}{p - \bar{k}_{_s}(\beta)}.
\]
One  can derive  from the above dispersion  equation (\ref{disp})
that the  resonances  are  analytic  functions  of the
perturbation  parameter $\beta$ near the  origin $\beta = 0$, but
the  factor ${\bf S}^{^{\beta}}_{_{0}} (p)$ is not for  the reason
discussed  above.

\begin{theorem} There  exist  a  one-dimensional perturbation
${\cal L}^{^{\beta}}_{_0}$  of  the  operator
\[
l_{_0} u =  -u'' ,\,\, u \bigg|_{_0} = 0
\]
with a  non-trivial  inner  component,  such that  the  scattering
matrix of the  pair  $({\cal L}^{^{\beta}}_{_0},\, l_{_0})$
coincides  with $-{\bf S}^{^{\beta}}_{_{0}} (p)$. Then the
Scattering matrix  ${\bf S}$ of the complementary pair
 $({\cal L}^{^{\beta}}_{_0},\,\,{\cal L}^{^{\beta}})$
is  equal to  the  complementary  factor $ - {\bf
S}_{_{\beta}}^{^{0}} (p) $:
\[
{\bf S}_{_{\beta}} (p) = {\bf S}^{^{\beta}}_{_{0}} (p)\,\,{\bf
S}_{_{\beta}}^{^{0}} (p).
\]
The  complementary  factor is an analytic function  of $(\beta,\,
k)$ on the product  of a  small  neighborhood of the origin  in
$\beta$-plane  and
 a  small  neighborhood  $K_{_0}$  of  $(k_{_{0}},\,-k_{_{0}} )$ in $p$-plane.
\end{theorem}
{\it Proof\,\,} Consider  the  one-dimensional  operator $A$ with
a positive eigenvalue  $\kappa^{^2}$. We associate with it the
Krein function ${\cal M} = \frac{1 - \kappa^{^{2}}\, \lambda
}{\kappa^{^{2}} - \lambda}$. Then restrict  the operator $l_{_0}$
onto smooth functions vanishing at the origin $x=0$, and construct
the self-adjoint extension ${\bf A}^{^{\beta}}_{_{0}}$ defined by
the hermitian matrix $\left\{{\bf {\beta}}\right\}$ which connects
the  boundary values  $\xi_{_{+}},\,- {\cal M}\xi_{_{+}} $ in  $
E_{_{A}}$ with boundary values $\Psi (0),\, \Psi'(0)$ in $
E_{_{+}}$:
\begin{equation}
\label{bcond0} \left(
\begin{array}{c}
\Psi'\\
-\xi_{_+}
\end{array}
\right) = \left(
\begin{array}{cc}
\beta_{_{00}}&\beta_{_{01}}\\
\beta_{_{10}}&\beta_{_{11}}
\end{array}
\right) \,\, \left(
\begin{array}{c}
\Psi\\
-\xi_{_-}
\end{array}
\right) .
\end{equation}
Substituting  into  the  above  equation the corresponding  Ansatz
for  the  wave-function  in the  outer space
\[
\Psi^{^{\beta}}_{_{0}} (x,p) = e^{- ipx}  +  {\bf S}^{^{\beta}}(p)
e^{- ipx},
\]
we  obtain  an  explicit expression  for  the corresponding
Scattering matrix  ${\bf S}$ in terms  of the inner Krein-function
${\cal M}(\lambda) = \frac{1 + \kappa^{^4} \lambda}{\kappa^{^2}-
\lambda}$ and  the  boundary  parameters:
\[
{\bf S} (\lambda) = \frac{ip + \left[ \beta_{_{00}} -
\frac{|\beta_{_{01}}|^{^2} {\cal M}}{1 + \beta_{_{11}} {\cal M}}
\right]} {ip - \left[ \beta_{_{00}} - \frac{|\beta_{_{01}}|^{^2}
{\cal M}}{1 + \beta_{_{11}} {\cal M}} \right]}.
\]
Select  the  parameter $\beta_{_{11}}$ such that $ 1-\beta_{_{11}}
\kappa^{^{2}} = 0, $ then  $1 + \beta_{_{11}} {\cal M}
=\beta_{_{11}} \frac{1 + \kappa^{^4}} {\kappa^{^2} - \lambda}$,
and  the  Scattering matrix takes  the  form  with $\lambda =
p^{^2}$
\[
{\bf S} (\lambda)= \frac{ ip + \beta_{_{00}} -
\frac{|\beta_{_{01}}|^{^2}}{\beta_{_{11}} (1 + \kappa^{^4})} -
\frac{|\beta_{_{01}}|^{^2} \kappa^{^{2}}}{(1 + \kappa^{^4})
\beta_{_{11}}} p^{^{2}}} { ip - \beta_{_{00}} +
\frac{|\beta_{_{01}}|^{^2}}{\beta_{_{11}} (1 + \kappa^{^4}) } +
\frac{|\beta_{_{01}}|^{^2} \kappa^{^{2}}}{(1 + \kappa^{^4})
\beta_{_{11}}} p^{^{2}}}
\]
Comparing  this  expression with
\[
-{\bf S}^{^{\beta}}_{_{0}} (p) = - \frac{[(p - k_{_0}(\beta)]\,[p
+ \bar{k}_{_0}(\beta)]}{[p - \bar{k}_{_0}(\beta)]\,[p +
k_{_0}(\beta)]} =
\]
\[
\frac{- p^{^{2}} + 2i {\Im k}_{_{0}}(\beta) p  + |
k_{_{0}}(\beta)|^{^{2}}}{ p^{^{2}} - 2i {\Im k}_{_{0}}(\beta) p  -
| k_{_{0}}(\beta)|^{^{2}}} = {\bf S} (\lambda),
\]
we  obtain the  equations  for  selection of parameters of the
extension  ${\bf A}^{^{\beta}}_{_{0}}$:
\begin{equation}
\label{eq1} \frac{\beta_{_{00}} -
\frac{|\beta_{_{01}}|^{^2}}{\beta_{_{11}} (1 + \kappa^{^{4}})}}
{|k{_{0}} (\beta)|^{^{2}}} = \frac{1}{2 \Im k_{_{_0}} (\beta)} =
\frac{|\beta_{_{01}}|^{^{2}} \kappa^{^{2}}}{(1 + \kappa^{^{4}})
\beta_{^{11}}}.
\end{equation}
 Solution is  not  unique, but  may be  parametrized  by  the
functional  parameter  $\beta_{_{00}} $ :
\[
\kappa^{^2} = \frac{1}{2 \beta_{_{00}} \Im k_{_{0}}(\beta) -
|k_{_{0}} (\beta)|^2}.
\]
 In particular we  may choose:
$\beta_{_{00}} = |k_{_{0}} (\beta)|^2 \left( \Im k_{_{0}}(\beta)
\right)^{^{-1}}$. Then
\[
\kappa^{^{2}} = \frac{1}{|k_{_{0}} (\beta)|^2 }.
\]
To  finalize  the  calculation  we  notice  that both $|k_{_{0}}
(\beta)|^2$  and  $\Im k_{_{0}}(\beta)$ are  real  analytic
functions  of  $\beta$, hence  $\kappa^2 $  is  an analytic
function of $\beta$. The  boundary  parameter  $\beta_{_{00}}$ has
a  pole  at  the  origin since $k_{_{0}} (\beta) \to  k_{_{0}} (0)
= \bar{k}_{_{0}} (0)$ when $\beta \to 0$. Other boundary
parameters may be also  chosen  based on above equations
(\ref{eq1}) as
 $\beta_{_{11}} = |k_{_{0}}|^{^2},\,\, |\beta_{_{01}}|^{^2} =
  \frac{4 + |k_{_0}|^{^4}}{4\Im k_{_0}}$. Alternative
choice  is  $\kappa^{^2} = |k_{_{0}}|^{^2},\,\, \beta_{_{00}} =
\frac{1 + |k_{_{0}}|^{^4}}{2\Im k_{_0}\,|k_{_0}|^{^2}},\,
|\beta_{_{01}}|^{^2} = \frac{1 + |k_{_{0}}|^{^4}}{2\Im
k_{_0}\,|k_{_0}|^{^4}},\,\, \beta_{_{11}} = |k_{_0}|^{^-2}$, with
$k_{_0} := k_{_0}(\beta)$. This accomplishes the construction of
the operator ${\bf A }^{^{\beta}}_{_{0}}$. \vskip0.5cm
 The operator ${\bf
A}^{^{\beta}}_{_0}$ permits  to  make a sort of a ``jump-start" in
perturbation procedure for the pair of operators $
\left\{\left(l\oplus H\right),\,{\bf A }_{_{\beta}}\right\} $: the
scattering matrix ${\bf S}^{^{\beta}}_{_{0}} (k)$ for  the pair  $
\left\{\left(l\oplus H\right),\,{\bf A}^{^{\beta}}_{_0} \right\} $
is calculated in explicit form, and the complementary scattering
matrix for  the pair $\left\{ {\bf A}^{^{\beta}}_{_0}, \, {\bf
A}_{_{\beta}} \right\}$, which
 coincides  with  ${\bf S}_{_{\beta}}^{^{0}}$, is an  analytic
 function of $(\beta,\, p) $ near to the  resonance  $k_{_0}$ and  can be calculated
via standard  analytical perturbation procedure.
\par
 Though the ``jump-start" can help  to  develop the  analytic
perturbation  procedure, according to  the preceding  result, but
exact  calculation  of the pole $k_{_{0}}(\beta)$  and
corresponding projection for the perturbed operator  ${\bf
A}_{_{\beta}}$, which were used in course of  construction of the
``jump-start", may be a tricky analytical problem, comparable with
the original spectral problem. On the other hand, the asymptotic
expansion for  these data  for small $\beta$ may be useful for
obtaining estimates, but  does not help to derive analyticity of
the complementary factor of the
 Scattering matrix constructed  for the  corresponding
 approximate ``jump start" factor, constructed  based on
 approximate  asymptotic data.
   \par
 \vskip1cm
\section{Conclusion: intermediate operator in analytic perturbation techniques}

Classical analytic perturbation techniques, see for instance
\cite{Kato}, is developed for additive perturbations of operators
with discrete spectrum $A \longrightarrow A + \varepsilon B$. For
perturbation of the continuous spectrum Kato suggested another
techniques based on Wave operators. Detailed study of the
perturbation of operators with continuous spectrum in model
situations, especially for the Friedrichs model, see
\cite{Friedrichs}, was done in numerous mathematical papers and
books, see \cite{simon} and the references therein and also
\cite{F64,PP70,FP84,H86}. It was already clear for specialists
long ago, that the classical techniques of the analytic
perturbation theory does not work generally on continuous spectrum
(actually it works below the threshold of creation of resonances,
see \cite{Friedrichs}). Probably I. Prigogine alone insisted on
modification of the analytic perturbation techniques to cover the
case of continuous spectrum.
    \par
According to evidence from long-time Prigogine's collaborator
Professor I. Antoniou, I.Prigogine attempted also to use the idea
of an {\it intermediate operator}  as a  base  for development of
the analytic perturbation techniques for operators with continuous
spectrum. He  tried  to construct the intermediate operator $
A_{_{1}}$ as a function of the non-perturbed  operator A with a
hope that after  the first  step  $A \to A_{_{1}}$  the analytical
perturbation procedure for  the  perturbed  operator  $ A +
\varepsilon V $ presented as  $ A_{_{1}}  +
  \varepsilon V'$  becomes  convergent. This  simplest  suggestion  appeared
to  be  not  so  efficient  as  expected,  and  Prigogine
abandoned this idea finally, see
\cite{Prig72,Prig73,Prig85,Antoniou,Prig89,Prig95}
\par
Note  that actually  the  idea of the intermediate  operator in
another context  appeared  in  {\it Splitting Method}  invented
 by I. Glazman, see \cite{Glazman}. This
method radically transformed  in  seventies the whole picture of
the qualitative spectral analysis  of singular differential
operators, but  was never used in context of the analytic
perturbation procedure.
\par
We actually suggested  revisiting
  the  idea of the  intermediate operator again in a special case of
quantum networks where the candidate for the role of an
intermediate operator is almost obvious. In our case it is not a
function of the  non-perturbed operators, but {\it is obtained
from it just by  a finite-dimensional perturbation}. Actually we
succeeded  to  apply the  splitting method  to  Quantum
 Networks because of  the  step-wise  structure
 of the  continuous spectrum in wave-guides. This  minor  change
 of  the  original idea  of  Prigogine  appears  to  be  useful
for  perturbation problems in wave-guides.  In our case the idea
of the intermediate operator is automatically combined with
resonances, which was anticipated  long  ago by Poincare and
intensely discussed  by Prigogine with references to Poincare, see
\cite{Prig89, Poincare}.
\par
The only essential, and even decisive difference of our version of
the intermediate operator from the version which  I. Prigogine
attempted to use, is the natural principle of {\it spectral
locality} we impose: we  construct the  intermediate operator and
the  scattered waves via the special perturbation procedure {\it
locally}- in a certain spectral interval, - but not globally, as
Prigogine planned. It is quite easy to define this interval
mathematically for the Quantum Networks: it is the maximal
interval containing the Fermi level, \cite{Madelung}, where the
spectral  multiplicity $n$ is constant and  the scattering  matrix
is a square matrix size $n \times n$. This interval may be either
a  spectral band corresponding to some channel in one of
semi-infinite  wires, of a  sum of several spectral  bands (sort
of a "spectral terrace") corresponding to different wires, with
properly hybridized scattered  waves. The structure of the
corresponding scattered waves within this spectral interval is
manifested by the structure of the scattering matrix and
corresponds  to  the n-dimensional input-output. The structure of
scattered waves on the neighboring spectral bands is different.
\par
It is  worth to  notice, that  slightly modified  procedure
permits  to  construct an  hierarchy of  intermediate  operators
also for some operators  with  absolutely-continuous spectrum of
constant  multiplicity, in particular  for  Friedrichs  model.
\par
In actual paper we  discussed  simplest  wave-guiding systems
composed  of  straight wave-guides (quantum wires). The
convenience  of this trivial environment appears from  existence
of explicit formulae for oscillatory modes in straight
wave-guides: the system of thresholds and the  structure  of  the
entrance sub-spaces in this environment is trivial. Study of  more
interesting systems of periodic or asymptotically straight
wave-guides with rapidly decreasing curvature requires a
combination of  our methods with Lippmann-Schwinger techniques
developed in case of wave-guides with non-trivial geometry for
discrete and especially for continuous spectrum, see
\cite{Twisted,Duclos,percylind,Krej_03, Krej6_03,Krej7_03}.
\par
On the other hand, the suggestion  to recover resonances from the
discrete spectrum of some intermediate  operator can be easily
applied  to any wave-guiding problem in trivial environment, with
linear vertices, on a reasonable price of preliminary  study of
the ``intermediate" dynamics. In particular, it permits to suggest
alternative approach  to  study of trapped modes in acoustics and
hydro-dynamics, investigated  in \cite{Evans,Davies}, as well as
to classical theory of electro-magnetic  wave-guides with local
variation of curvature, \cite{DES}. An important  advantage of our
techniques is: the possibility of  ``fitting'' of the constructed
``solvable model'' based  of  results of straightforward computing
for the intermediate  problem.

\section{Acknowledgement }
Two of co-authors (A.M and B.P) acknowledge  support from the
Russian Academy of Sciences, Grant RFBR  03-01- 00090 and from
research allocation of  the  University of Auckland. All authors
are grateful to Doctors M. Harmer, V. Olejnik and V. Kruglov for
interesting comments on the text and to  Dr. R. Shterenberg for
extended discussion of questions concerning the singular spectrum
in channels, important references and materials \cite{KNP}
supplied. The  authors are grateful to Professor I. Antoniou for
inspiring information on Prigogine's experience in analytic
perturbation theory and interesting materials  supplied.
\par
BP is grateful to Professor L. Faddeev for an extended discussion
of the perturbation of the  continuous  spectrum  for  the
Friedrichs model, which  clarified  the  structure  of the
hierarchy of intermediate  operators.

\end{document}